# EXPLORING FUNDAMENTAL PHYSICS WITH NEUTRON STARS

PIERRE M. PIZZOCHERO

*Dipartimento di Fisica, Università degli Studi di Milano, and
Istituto Nazionale di Fisica Nucleare, sezione di Milano,
Via Celoria 16, 20133 Milano, Italy*

## ABSTRACT

*In this lecture, we give a first introduction to neutron stars, based on fundamental physical principles. After outlining their outstanding macroscopic properties, as obtained from observations, we infer the extreme conditions of matter in their interiors. We then describe two crucial physical phenomena which characterize compact stars, namely the gravitational stability of strongly degenerate matter and the neutronization of nuclear matter with increasing density, and explain how the formation and properties of neutron stars are a direct consequence of the extreme compression of matter under strong gravity. Finally, we describe how multi-wavelength observations of different external macroscopic features (e.g. maximum mass, surface temperature, pulsar glitches) can give invaluable information about the exotic internal microscopic scenario: super-dense, isospin-asymmetric, superfluid, bulk hadronic matter (probably deconfined in the most central regions) which can be found nowhere else in the Universe. Indeed, neutrons stars represent a unique probe to study the properties of the low-temperature, high-density sector of the QCD phase diagram. Moreover, binary systems of compact stars allow to make extremely precise measurements of the properties of curved space-time in the strong field regime, as well as being efficient sources of gravitational waves.*

## I – INTRODUCTION

Neutron stars are probably the most exotic objects in the Universe: indeed, they present extreme and quite unique properties both in their macrophysics, controlled by the long-range gravitational and electromagnetic interactions, and in their microphysics, controlled by the short-range weak and strong nuclear interactions. Actually, neutron stars are as astounding to the astrophysicist as they are to the nuclear physicist: what the former sees as an incredibly compact star appears to the latter as an incredibly extended nucleus, compressed and bound by gravity. Their extreme stellar properties (e.g., gravitational field, rotational frequency, magnetic field, surface temperature) are well matched by their extreme nuclear properties (e.g., nucleon density, isospin asymmetry, bulk superfluidity and superconductivity, lepton cooling by neutrinos).

A comprehensive study of neutron stars requires an understanding of how the long and short range interactions affect each other. Indeed, the physics of compact stars and of the stellar systems they form constitutes a new and thriving field of research, *relativistic nuclear astrophysics* (also called *astronuclear* physics), which requires expertise from disciplines that are generally mostly disjointed, but now have to work sinergically: high-energy astrophysics, gravitational physics (i.e., General Relativity), nuclear and hadronic physics, neutrino physics, QCD, superfluid hydrodynamics, plasma physics. Therefore, presenting the subject of neutron stars to students and early-stage researchers is not a simple task: the interdisciplinary nature of this study makes it a long, although rewarding, process. Several books and many reviews on the different topics involved are presently available [1] and regular schools are organized every year. In Europe, the network *Compstar* [2] is dedicated to the physics of compact stars and to the formation of young astronuclear physicists.

Obviously, the goal of this lecture is not to present a complete and detailed overview of neutron stars, of their properties and of the physical processes involved: on top of being impossible in such a short lecture format, this would be useless to an audience of nuclear and sub-nuclear physicists, who are trained in the physics of microscopic systems of degenerate nucleons or high-energy particles, but are probably not familiar with astrophysical issues in general, and with the macroscopic behaviour of objects strongly bound by gravity in particular. Here, instead, we introduce the subject of neutron stars by first showing how some of their observed macroscopic properties, and the internal physical conditions we infer from them, are easily explained as a consequence of the strong gravitational field which, in this final stage of stellar evolution, compresses matter to its very limit; actually, it is the dominant effect of gravity on the other interactions which characterizes the physics of compact stars and the stellar systems they form. We then explain why observations of neutron stars (isolated or in binary systems) can be invaluable to fundamental physics; on the one hand, they provide a unique astrophysical laboratory to test the properties of hadronic matter under extreme conditions, not attainable in any terrestrial facility now or in the future. On the other hand, they allow the most stringent available tests on the properties of space-time and thence on general relativity, at present the most successful theory of gravity.

The lecture develops as a series of answers to four basic questions. In Section II, we introduce the main observed properties of neutron stars and deduce from them their internal physical conditions. In Section III, we briefly describe the gravitational stability of degenerate fermions and the neutronization of matter during gravitational contraction, with which we can explain the origin of neutron stars, their mass range and the conclusion that these stars are basically macroscopic assemblies of neutrons bound by gravity. In Section IV, we outline the composition of hadronic matter along the star profile, from the superfluid crust to the super-dense exotic core, as determined by the increasing density with depth. In Section V, we discuss how the study of neutron stars can be used to address some fundamental physical issues. In particular, we choose as examples three significant classes of observations (maximum mass, surface cooling and pulsar glitches) and we outline how these observations can yield invaluable information about some crucial questions in hadronic physics: the equation of state (EOS) of dense bulk matter, macroscopic nucleon superfluidity, and the properties of neutron-rich exotic nuclei above neutron-drip. Moreover, we briefly discuss the unique opportunity presented by binary systems of compact stars to precisely measure masses through general relativistic effects, as well as being strong sources of gravitational waves.

## II – WHAT ARE NEUTRON STARS?

Although predicted theoretically soon after the discovery of the neutron in 1932, neutron stars were first observed accidentally in 1967 in the radio band, as periodically pulsating sources called *pulsars*; since then, about two thousands have been discovered. These radio-pulsars have periods in the range $P \sim 10^{-3} - 10\ s$, and although they gradually slow down with period derivatives in the range $\dot{P} \sim 10^{-20} - 10^{-10}\ s\ s^{-1}$ (see Figure 1a), their periods are so stable that pulsars can be considered the most accurate clocks in the Universe. Coupled to independent estimates of the mass (see later), a simple analysis of the pulsed emission implies that pulsars must be *rotating compact objects* with mass of the order of that of the Sun (the solar mass is $M_\odot = 2 \times 10^{33}\ g$) and radius of the order of $R = 10^6\ cm$. Indeed, among periodic mechanisms, stellar pulsations cannot explain at the same time the slow and fast pulsars, while binary systems losing energy by emission of gravitational waves would spin up instead of slowing down. This leaves only rotations, but the shortest periods require very compact objects: a $1\,M_\odot$ white dwarf (typical radius $\sim 10^4\,km$) with a period of one millisecond would be above the keplerian limit for the maximum angular frequency $\Omega_k = 2\pi/P_k = \sqrt{GM/R^3}$ (if $\Omega > \Omega_k$, the gravitational field is not strong enough to provide the centripetal acceleration) and would fly apart; a black hole, instead, has no surface and thence cannot emit periodically. The identification of these compact objects of mass $\sim 1\,M_\odot$ and radius $\sim 10\,km$ with stars made of *neutron* will be explained in Section III.

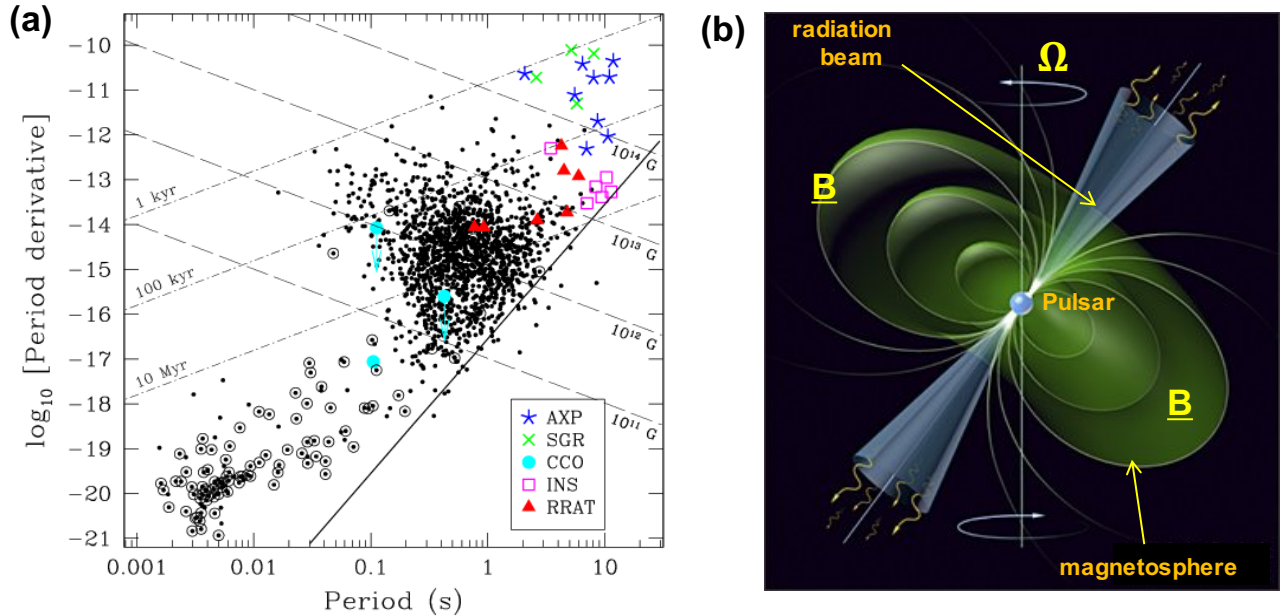

*Figure 1 – (a) $P$-$\dot{P}$ diagram for observed neutron stars. Lines of constant age and constant magnetic field are drawn, according to the rotating dipole model. Black dots are RPPs, circled dots are RPPs in binary systems, AXP and SGR are magnetars, INS and CCO are thermal emitters. – (b) The lighthouse model for pulsar periodic emission from a rotating magnetic dipole.*

The emission from radio-pulsars was soon interpreted in terms of rotating neutron stars with large dipolar magnetic fields: the energy emitted by a rotating (i.e., accelerated) oblique magnetic dipole in vacuum would slow down the pulsar and explain the steadily increasing period. This is the so-called "lighthouse model" (see Figure 1b), where the rotation of a narrow beam of radiation is seen by the distant observer (if in the right line of sight) as a pulsation; it was later supported by the evidence of cyclotron lines in the spectra, corresponding to huge magnetic fields of order $B \sim 10^{12}\,G$. The dipole model determines the age, $\tau$, and magnetic field of a pulsar from the measured values of $P$ and $\dot{P}$; for example, $\tau = P/2\dot{P}$ (lines of constant $\tau$ and $B$ are shown in Figure 1b). Although model dependent, observations yield the ranges $\tau \sim 10^3 - 10^{10}\,yr$ and $B \sim 10^7 - 10^{15}\,G$, with some of the youngest objects having the largest magnetic fields (*magnetars*) and the oldest ones having the fastest rotation (*millisecond pulsars*); the large majority of pulsars, however, can be found in the range $\tau \sim 10^5 - 10^9\,yr$ and $B \sim 10^{11} - 10^{13}\,G$. These different classes of stars are generally explained in terms of stellar evolution; for example, millisecond pulsars, which are mostly found in binary systems, are well interpreted as old neutron stars with a decayed magnetic field that have been spun up by accretion of mass from the companion star and thence recycled as fast pulsars. Magnetars, on the other hand, appear to be young neutron stars which, for reasons that are still under study, were born with extremely large magnetic fields (in excess of the QED critical field $B_{QED} = 4.4 \times 10^{13}\,G$).

It is worth stressing that the oblique magnetic dipole rotating in vacuum is an extremely oversimplified model of pulsar emission; indeed, it turns out that the huge electric fields induced by rotation will pull charged particles away from the surface, so that the neutron star is actually surrounded by a complex magnetosphere filled with charges of both signs strongly coupled to the electromagnetic field. Assuming only magnetic dipole breaking, however, the dependence of the total emitted power on the star macroscopic properties (spin, radius and surface magnetic field) turns out to be the same as in the simple vacuum model, namely $dE_{dip}/dt \propto \Omega^4 B^2 R^6$ where $\Omega = 2\pi/P$ is the pulsation. This simple dipole relation is normally used to describe *Rotation-Powered-Pulsars* (RPP), whose source of energy is their rotational kinetic energy (the emission from magnetars, instead, is driven by their magnetic energy) and which represent the vast majority of the objects observed so far and shown in Figure 1a.

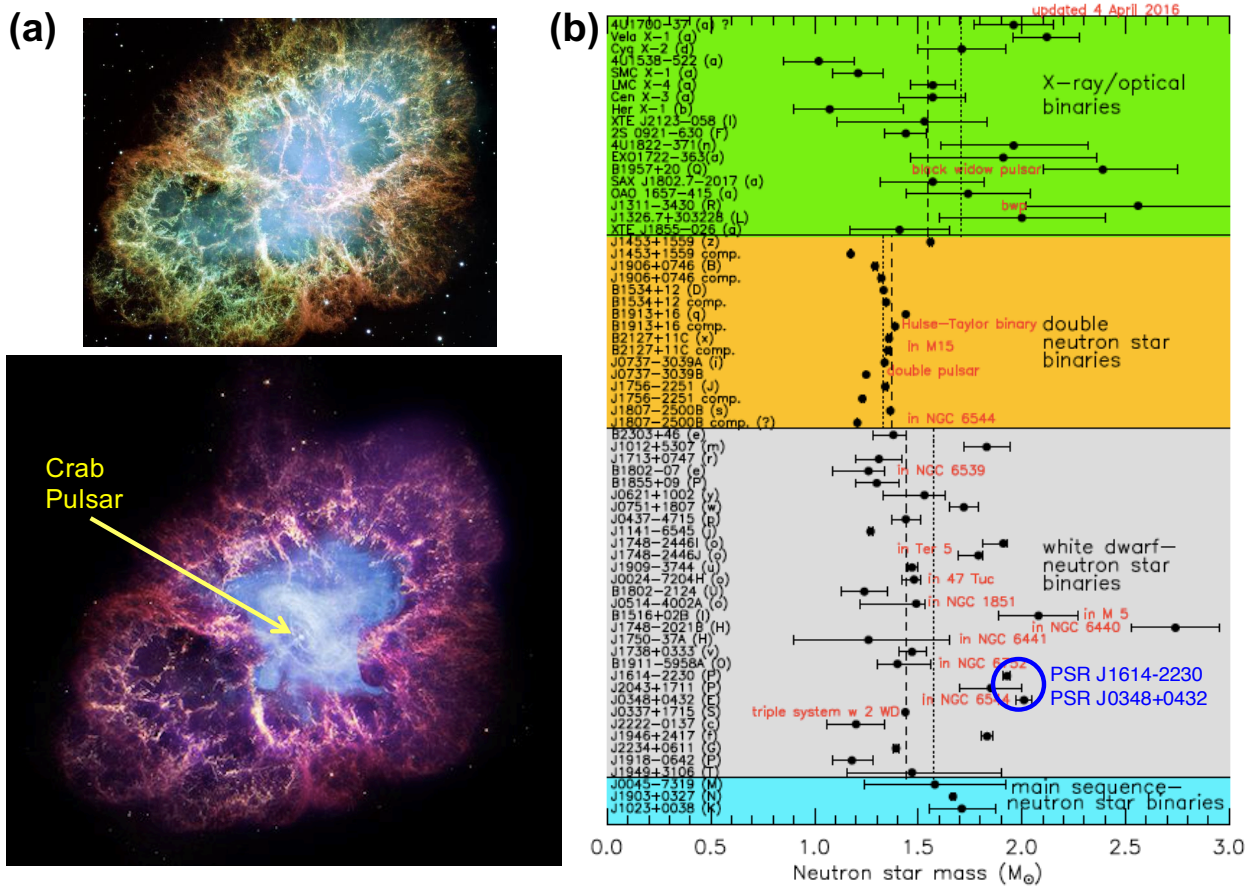

*Figure 2 – (a) The Crab Nebula (top photo, optical band) is the remnant of a Supernova exploded in 1054. At present, it is powered by the Crab Pulsar (bottom photo, composite optical, IR and X bands), whose emission of radiation and particles interact with the ejecta forming the central Pulsar Wind Nebula (observable only in X-rays). – (b) Measured masses of neutron stars in binary systems; the data are divided according to the nature of the companion star in the system (from Ref. [3-b]). The recent precise measurements of massive ( ~$2M_\odot$ ) neutron stars are highlighted.*

The actual emission from RPPs is not expected to be simple dipole radiation, but rather a complex interplay of low-frequency radiation and a wind of relativistic accelerated particles, in general not directly observable for isolated pulsars but only indirectly through their secular spin-down ($\dot{\Omega} < 0$). For young neutron stars still surrounded by the debris of the Supernova which originated them, however, this emission can interact with the the surrounding ejecta creating a *Pulsar Wind Nebula*, whose complex structure can be revealed only in the high energy band (as shown in Figure 2a for the Crab system). As for the *pulsed* radiation actually *observed* from RPPs, which can extend over the *entire* electromagnetic spectrum from radio frequencies up to $\gamma$-rays, it constitutes a minimal fraction of the total energy loss (as measured by the pulsar spin-down $dE_{rot}/dt \propto MR^2\Omega\dot{\Omega}$), and it originates in different parts of the magnetosphere and at different heights from the star, through relativistic processes of particle acceleration and pair cascades which emit collimated beams of radiation. This is a fascinating and active field of research which applies radiation and plasma physics to the extreme conditions found in neutron star magnetospheres.

Like for all stars, neutron star masses can be measured when they belong to binary systems, using Kepler's third law coupled to other orbital data. In particular, double neutron star binaries present additional observable general relativistic effects (post-Keplerian parameters), due to the very strong gravitational field, that allow a very precise determination of the mass [3-e]. Figure 2b reveals a mass range $M \sim 1 - 2\ M_\odot$, with an error-weighted mean value (dashed vertical line) of $M \simeq 1.4\ M_\odot$. The best determined masses used to lie in a narrow interval $M \sim 1.25 - 1.45\ M_\odot$, while values $M \gtrsim 2\ M_\odot$ were affected by large error bars. Recently, however, very precise measurements have revealed two massive pulsars with $M \cong 2\ M_\odot$ (highlighted in Figure 2b); later in the lecture we will come back to the relevance of this for fundamental physics.

Neutron star radii can only be determined indirectly. For example, if the distance of the star from Earth is known, its radius can be obtained by fitting model atmospheres to the observed spectra; or, if the mass is known, it can be inferred by the gravitational redshift of spectral lines. Although very model-dependent, the estimated radii lie in a range $R \sim 10 - 15\ km$, consistent with the keplerian limit discussed earlier. Better atmosphere models have been developed recently and the best constraints from observations now give radii in the range $R \sim 11.5 - 12.5\ km$, but a reliable determination of the radius is still an open issue.

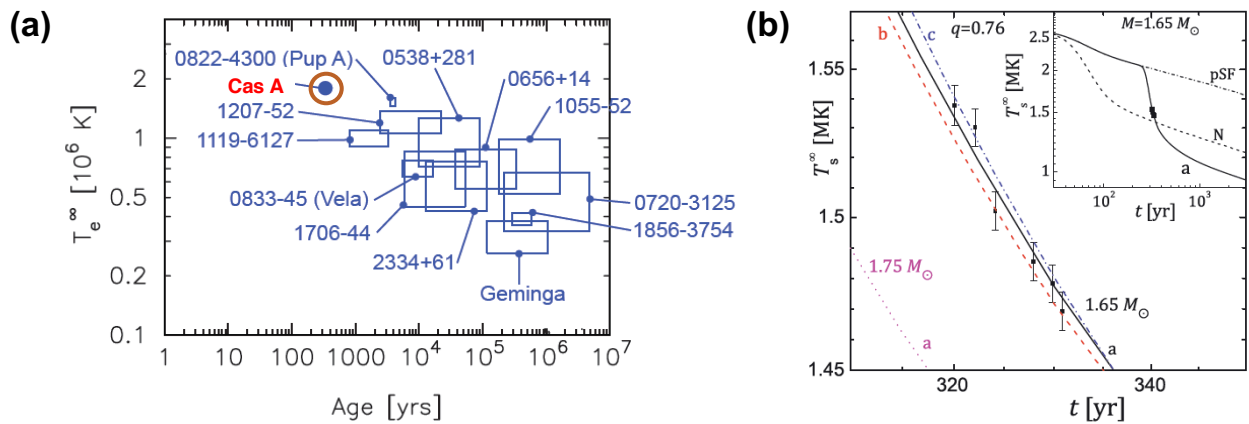

*Figure 3 – (a) Measured surface temperatures of isolated neutron stars; the ages reported in abscissa are obtained from the rotating dipole model. Notice the large error boxes, with the notable exception of Cassiopeia A. – (b) Cooling of the CCO in the Cassiopeia A supernova remnant. The six points represent measurements taken by the Chandra X-ray satellite over the past decade and the lines are theoretical cooling curves (from Ref. [3-d]).*

Surface temperatures can be measured for several isolated neutron stars of different ages (cf. Figure 3a), whose thermal emission is not hidden by strong non-thermal magnetospheric emission, like it is usually the case. In particular, the so called *X-ray-dim Isolated Neutron Stars* (labelled INS in Figure 1a) present pure blackbody spectra, consistent with thermal emission from the stellar surface. We point out the large error boxes in the results shown in Figure 3a: indeed, fitting the observed spectra (e.g., with a black-body or with a model atmosphere) and extracting an effective surface temperature and luminosity is not a model-independent procedure, as already discussed for the radius measurements. Similarly, the age of the pulsar is derived from the rotating dipole model, which is obviously only a first approximation to the complex emission mechanism, associated to the extended magnetosphere of charged particles strongly coupled to the magnetic field, which in turn is anchored to the spinning neutron star. In spite of the uncertainties, neutron stars appear to cool down as they age, their surface temperatures spanning the interval $T_s \sim 3\times 10^5 - 2\times 10^6\ K$, namely they shine in the soft X-ray range. A remarkable source is the *Compact Central Object* (CCO) detected in Cassiopeia A, the youngest and warmest neutron star observed so far: its age is well known, since it is associated to the remnant of a Supernova which exploded around 1680 AD. Its surface temperature not only is well determined, but in the last decade it has been observed to decrease with time [3-d], as shown in Figure 3b. To date, this is the only direct evidence of cooling of a neutron star; later in the lecture we will discuss its relevance for fundamental physics.

Incidentally, these temperature observations can only be made from telescopes mounted on satellites, since the Earth's atmosphere is opaque to most electromagnetic waves outside the visible range, except for a large transparency window in the radio frequency range, a fact which made the early discovery of (radio-)pulsars possible. Nowadays, thanks to the existing fleet of space telescopes, pulsars are observed in all the different bands of the electromagnetic spectrum and their pulsed emission can also be detected at the IR, visible, UV, X and γ energies, in several possible combinations: the Crab and Vela pulsars, for example, pulse in all bands, Geminga is a γ-ray pulsar, the CCOs and the INSs are radio-quiet and so on. We are in the presence of a *zoo* of neutron stars, broadly distinguished by the source that power they emission (for a review, see Ref. [3-c]): rotational kinetic energy (RPPs), thermal energy (INSs and CCOs), magnetic field energy (magnetars), accretion from a companion star (X-ray binary pulsars). Although the normal RPPs constitute to date the majority of the population, this could be an observational bias; indeed, in the last two decades new satellites have been changing our view about neutron stars: for example, the *Fermi* satellite has discovered a large new population of γ-ray pulsars, while only Geminga was known before its launch. With a new generation of telescopes under way, multi-wavelength observations of the sky will surely reveal unknown facets of compact stars.

We thus see how observations of several properties of pulsars point to quite extraordinary objects from the astrophysical point of view: stars with mass similar to that of the Sun but compressed by gravity into a spherical volume 20-30 km in diameter, spinning around their axis up to a thousand times per second, shining in the soft X-ray, with a powerful magnetosphere generated by magnetic fields more than a billion times stronger than those in the Sun. This magnetosphere, which is like an extension of the neutron star itself, being dragged along by the rapidly rotating star and magnetically containing intense flows of relativistic charged particles, is responsible for both the observed pulsed emission and the much larger loss of energy by magnetic dipole radiation and particle acceleration (pulsar wind). If the star has a mass quadrupole moment, energy can also be lost by emission of gravitational waves. For the rotationally-powered pulsars we are considering here, the bulk loss of energy is made evident by the star's slow-down. In the case of the Crab pulsar, the evidence is even more direct: found in the middle of the very luminous Crab Nebula, which is the remnant of a supernova exploded in 1054 and observed by Chinese astronomers, the Crab pulsar powers the Nebula which surrounds it and maintains it very bright even a thousand years after the supernova explosion. Indeed, the decrease in rotational kinetic energy of the Crab pulsar, $dE_{rot}/dt = I\Omega\dot\Omega$ (within a factor of order unity, the momentum of inertia is $I \simeq MR^2 \sim 10^{45}\ g\ cm^2$, when evaluated for typical neutron star parameters), matches the total luminosity of the Crab Nebula ($L_{\text{Crab}} \sim 5\times 10^{38}\ erg\ s^{-1}$).

To understand why neutron stars are extraordinary also from the microscopic point of view, we must infer their unseen internal properties from the observed astrophysical parameters, and thence determine the physical state of matter in their interiors. The *average* mass density corresponding to $M = 1.4\ M_\odot$ and $R = 15\ km$ is $\bar\rho = 3M/4\pi R^3 \simeq 2\times 10^{14}\ g\ cm^{-3}$; this is almost the value of nuclear saturation density $\rho_0 = 2.8\times 10^{14}\ g\ cm^{-3}$, namely the density of ordinary atomic nuclei; stars with $R = 10\ km$ would have $\bar\rho \simeq 2.3\ \rho_0$. Due to the steep gradients in pressure and density necessary to stabilize the star against gravity, we expect much larger central densities $\rho_c$: for example, the simple polytropic model (a first approximation to represent gravitationally bound spheres of degenerate fermions) predicts $\rho_c = 6\bar{\rho}$ for non-relativistic particles and $\rho_c = 54\bar{\rho}$ for ultra-relativistic ones. We are thus dealing with *super-dense* matter, namely nucleons compressed by gravity well beyond the average density of normal nuclei. Nowhere else in the present Universe can matter with such properties be found and observed: we have the unique scenario of extended (bulk) compressed nuclear matter. In Section III we will argue that it is also strongly isospin-asymmetric, being mostly composed of neutrons (less than five protons every hundred neutrons). Since a solar mass of matter contains $N_\odot = M_\odot/m_N \simeq 10^{57}$ nucleons (where $m_N$ is the nucleon mass), neutrons stars could also be viewed as giant kilometer-sized nuclei with $A \simeq 10^{57}$, $Y_e = Z/A \lesssim 0.05$ (in astrophysics, the ratio $Z/A$ is indicated by $Y_e$ since it also represents the number fraction of electrons with respect to nucleons) and radius $R_A = r_0 A^{1/3} \simeq 10^6 \, cm$ (with $r_0 = 1.2 \, fm$). Such a point of view is somewhat misleading, however, since these objects are held together by gravity alone and *not* by nuclear forces (which, incidentally, are not even able to bind two neutrons).

To further study the state of matter inside neutron stars let us consider a density $\rho = 2\rho_0$ (but keep in mind that this is quite conservative and that in the central regions even larger densities are expected). The number density of nucleons is $n = \rho/m_N = 2n_0 \simeq 0.34 \, fm^{-3}$ (where $n_0 = \rho_0/m_N = 0.17 \, fm^{-3}$ is the nucleon density at saturation) and thence their average distance is $\bar{l} \sim n^{-1/3} \sim 1.4 \, fm$; on the other hand, the thermal De Broglie wavelength of the nucleons is $\lambda_T \sim h/\sqrt{m_N k_B T} \sim 4 \times 10^6 \, T^{-1/2} \, fm$. The temperatures inside neutron stars can be inferred from cooling calculations (see Section V) to reach $T \lesssim 10^9 \, K$ shortly (a few days) after the star birth, independent from the initial temperature of formation (which is estimated of order $10^{11} - 10^{12} \, K$), so that it results $\lambda_T > 10^2 \, fm$. Therefore, in spite of very high temperatures, up to a billion degrees for very young neutron stars (for comparison, the interior of the Sun has $T_\odot \sim 10^7 \, K$), due to the extremely large densities we obtain the strong inequality $\lambda_T \gg \bar{l}$ in neutron stars interiors: this indicates that a classical description of particles is not valid in such a regime, since their quantum wave-like properties cannot be neglected at the typical inter-particle scale. In these conditions, gravity has compressed the nucleons so much that they are even closer than they usually are under the action of nuclear forces alone; they are squeezed to occupy the lowest available energy states according to the Fermi-Dirac quantum distribution for fermions. The temperature, although very large, is not anymore a measure of the average particle energy; the degeneracy (zero-point) energy, which is due to the Pauli exclusion principle and is density-dependent, now completely dominates the thermal contribution. In conclusion, baryonic matter in neutron stars is *strongly degenerate* and a zero-temperature approximation is physically meaningful.

To proceed, we take the Fermi model at $T = 0$ as a guide, namely we neglect interactions and assume that the nucleons occupy all the lowest momentum states available in phase space up to a maximum value, the Fermi momentum $p_F$. We then have the standard result $p_F = \hbar k_F$, with the Fermi wavenumber $k_F = (3\pi^2 n)^{1/3} \simeq 1.7 \, (n/n_0)^{1/3} \, fm^{-1}$. For non-relativistic particles (see later on), we have the density-dependent Fermi energy $E_F = p_F^2/2m_N \simeq 59 \, (n/n_0)^{2/3} \, MeV$. The condition $\lambda_T \gg \bar{l}$ is equivalent to $E_F \gg k_B T$, as expected for strongly degenerate particles; indeed, for $n = 2n_0$ we find $p_F \simeq 430 \, MeV/c$ and $E_F \simeq 94 \, MeV$, while $T < 10^9 \, K$ implies $k_B T < 0.1 \, MeV$. The relativity parameter for the nucleons is $\xi = p_F/m_N c \simeq 0.3 \, (n/n_0)^{1/3}$; for $n = 2n_0$ we have $\xi \sim 0.4$, and for *any* reasonable density we still find $\xi \lesssim 1$; therefore, nucleons inside neutron stars are partially relativistic: they are neither ultra-relativistic ($\xi \gg 1$) nor completely non-relativistic ($\xi \ll 1$).

In conclusion, the observed compactness of neutron stars implies that at the microscopic level they do not behave as classical systems (like normal extended stars, e.g. the Sun), but rather they present extreme quantum properties where temperature can be neglected and the internal energy depends on particle density. Matter resists compression by a zero-point pressure which is a direct consequence of the Pauli exclusion principle, acting as an effective repulsion between nucleons.

The gravitational stability of stars supported by the pressure of completely degenerate, partially relativistic fermions will be discussed in the next section.

## III – WHY STARS MADE OF NEUTRONS?

In order to understand why the compact objects that are observed are made of neutrons, to follow in its main steps the evolutionary process that forms neutron stars and to explain the limited range observed for their masses, we must develop two physical ideas, one macroscopic and the other microscopic, but both related to the presence of a strong gravitational field in compact degenerate stars. The first idea is that the degeneracy pressure of fermions is only able to resist gravity up to a *maximum stellar mass*, called Chandrasekhar's mass. If the star's mass is larger than this limit, the star is not gravitationally stable and it must collapse to a different configuration (in the case of neutron stars, a black hole). The second idea is that above densities of order $\rho_\beta \sim 10^7 \; g \; cm^{-3}$, increasing the matter density induces electron capture on nuclei and free protons and prevents neutron β-decay, namely above $\rho_\beta$ we have an increasing degree of *neutronization of matter*.

The existence of maximum allowed masses for different classes of stars is related to a very general property derived from the gravitational virial theorem: a self-gravitating object is bound and stable as long as the particles that provide the pressure to resist gravity are non-relativistic; as the particle energies approach relativistic values, because of increasing temperature or density, the system becomes gravitationally unstable to small perturbations and, for ultra-relativistic particles, it is no longer bound together by gravity. This can be applied to different scenarios; for example, classical (extended) stars tend to become hotter, and thus increasingly dominated by radiation (i.e., ultra-relativistic) pressure, the more massive they are; indeed, stars above $50 \, M_\odot$ are very rare and those above $100 \, M_\odot$ are exceptions. In the present case of degenerate fermions in compact stars, the gravitational instability and the maximum mass are also related to the pressure-providing particles (electrons for white dwarfs, nucleons for neutron stars) becoming relativistic. To better explain the origin of the Chandrasekhar's mass limit for compact stars, the following argument, due to Landau, is as simple as it is revealing.

We consider a star of mass $M$ and radius $R$, made of $N = M/m$ completely degenerate ($T = 0$) fermions of mass $m$. The (average) number density of particles is $n \simeq N/R^3$ and their Fermi momentum is $p_F \simeq \hbar \, n^{1/3} \simeq \hbar \, N^{1/3} R^{-1}$. The total gravitational energy of the star is $E_G \simeq -G \, M^2/R$ (we neglect factors of order unity; for a uniform-density sphere we would have a factor of $3/5$) and thus the gravitational energy per particle is $U \simeq -Gm^2 \, N/R$. The (average) kinetic energy of the particles is just their Fermi energy, namely $K \simeq E_F$ (where again we omit factors of order unity, in the present case $3/5$ or $3/4$ for non-relativistic or ultra-relativistic particles) where $E_F = p_F^2/2m \simeq \hbar^2/m \; N^{2/3} R^{-2}$ for $\xi \ll 1$ and $E_F = p_F c \simeq \hbar c \, N^{1/3} R^{-1}$ for $\xi \gg 1$. The total energy per fermion is then $E = U + K$ and its behaviour with $R$ determines the stability properties of the star. In Figure 3, we plot $E$ as a function of the star's radius. For large values of $R$, the density is low and the particles are non-relativistic ($\xi \ll 1$), so that $E \simeq -Gm^2 \, N/R + \hbar^2/m \; N^{2/3} R^{-2} \approx -1/R$ as $R \to \infty$. For small values of $R$, the density is high and the particles are ultra-relativistic ($\xi \gg 1$) so that $E \simeq -Gm^2 \, N/R + \hbar c \, N^{1/3} R^{-1} \approx (-Gm^2 \, N + \hbar c \, N^{1/3})/R$ as $R \to 0$. Because of the term in parenthesis, there is a critical value of the total particle number $N_{\max} \simeq (\hbar c/Gm^2)^{3/2}$ and of the star's mass $M_{\max} \simeq m(\hbar c/Gm^2)^{3/2}$, such that (as $R \to 0$) $E \approx +1/R$ if $M < M_{\max}$ and $E \approx -1/R$ if $M > M_{\max}$. The asymptotic behavior just described allows to sketch the *shape* of the curves $E = E(R)$, as shown in Figure 3, where the solid line corresponds to masses $M < M_{\max}$ and the dashed line to $M > M_{\max}$.

The minimum in the curve for $M < M_{\max}$ shows that such stars have a stable equilibrium configuration for some value $R_0$ of the radius; conversely, stars with $M > M_{\max}$ cannot be stabilized against gravity by the pressure of degenerate fermions alone and are bound to collapse. Taking $m = m_N$ as it is the case for neutron stars, where nucleons provide both mass and pressure, we find $N_{max} \simeq 2\times 10^{57}$ and $M_{max} \simeq 1.5\, M_\odot$. The order of magnitude of the equilibrium radius for stars with $M \lesssim M_{max}$ can be estimated by noting that the departure of the solid curve from the $-1/R$ behavior as $R$ decreases is due to the particles becoming relativistic, so that $R_0$ must be small enough to obtain $\xi \gtrsim 1$. Since $\xi = p_F/m_N c \simeq \hbar/m_N c\; N^{1/3} R^{-1}$, the value of $R_0$ must satisfy $R_0 \lesssim \hbar/m_N c\; N_{max}^{1/3} \simeq 20\, km$. It is quite amazing how the macroscopic properties of neutron stars appear naturally from the microscopic behavior of their constituent particles: their scale is determined by simple combinations of fundamental constants (e.g., $N_{max} \simeq \alpha_G^{-3/2}$, where $\alpha_G = G m_N^2/\hbar c \simeq 6\times 10^{-39}$ is the gravitational equivalent of the fine-structure constant $\alpha = e^2/\hbar c \simeq 1/137$ of electromagnetism, and sets the strength of gravity in baryonic matter).

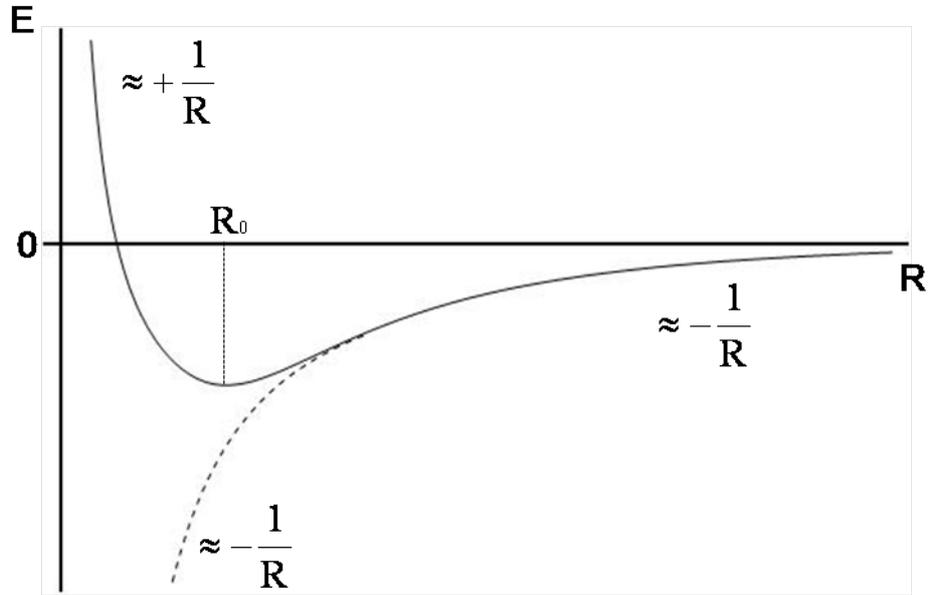

*Figure 3 – Energy per particle in a star of mass M and radius R, whose pressure is provided by N completely (T = 0) degenerate fermions. The asymptotic behavior of the energy is indicated for large (non-relativistic regime) and small (ultra-relativistic regime) values of the radius. The solid curve corresponds to $N \lesssim 10^{57}$, and has a stable bound state at $R = R_0$. The dashed line corresponds to $N > 10^{57}$ and leads the star to a complete collapse.*

Of course, the previous argument only sets the scale of the maximum mass of neutron stars, not its precise value; in particular, we have neglected the strong interaction between nucleons, which contributes to the internal energy and significantly affects the maximum mass; we will come back to this issue in Section V. For now, we just mention the existence of a strong upper limit to the mass of a stable neutron star, $M \lesssim 3\, M_\odot$, derived from General Relativity under very reasonable assumptions (microscopic stability and causality of the EOS of dense matter above saturation): this is the main observational criterion used to discriminate black holes from neutron stars.

Landau's argument could be extended to describe *white dwarfs,* namely inert stellar objects stabilized against gravity by the degeneracy pressure of their *electrons*, and which appear as final stages of stellar evolution (see later on). The argument would yield the same maximum mass, but a planet-like radius of order $\sim 10^4\, km$ (since now it is the electrons rather than the nucleons that become relativistic) and thence average densities of order $\sim 10^6\, g\, cm^{-3}$. At these densities, matter is described by classical (thermal) ions and strongly degenerate electrons, the latter providing most of the pressure. White dwarfs were first studied theoretically by Chandrasekhar; in the limit $T = 0$ (complete degeneracy) and for non-interacting electrons, he calculated their limiting mass as $M_{\text{Ch}} = 5.76\,Y_e^2\,M_\odot$, where $Y_e = Z/A$ is the electron fraction corresponding to ions of charge $Z$ and mass number $A$. Typically, $Y_e \simeq 1/2$ so that $M_{\text{Ch}} \simeq 1.44\,M_\odot$; we will come back to this value at the end of this section.

We now turn to discuss the neutronization of matter with increasing density. This is a direct consequence of the shift in β-equilibrium induced by the presence of increasingly relativistic degenerate electrons, whose total energy grows with density much faster than that of the non-relativistic nucleons. To understand the physical principle, we take the simplest case of capture on protons and study the equilibrium of the weak process $p + e \leftrightarrows n + \nu$ (capture on nuclei will be discussed in Section IV). The thermodynamical equilibrium of protons, neutrons, electrons and neutrinos is expressed by the equality $\mu_p + \mu_e = \mu_n + \mu_\nu$ between their chemical potentials, coupled to the charge-neutrality condition $n_p = n_e$. As explained before, we now take completely degenerate non-interacting particles at $T = 0$; we then have $\mu_i = E_{F,i} = \sqrt{(m_i c^2)^2 + (p_{F,i} c)^2}$ and $p_{F,i} = \hbar\,(3\pi^2 n_i)^{1/3}$ ($i = n, p, e, \nu$), with $m_i$ the respective rest masses. The complete relativistic expression for the energy *must* be taken, when particle reactions occur and the mass-energy equivalence is crucial in the total energy balance. Since the massless neutrinos escape from the system ($n_\nu = 0$), we take accordingly $\mu_\nu = 0$ and the β-equilibrium condition becomes $E_p + E_e = E_n$ (where we simplify the notation as $E_i \equiv E_{F,i}$ and $p_i \equiv p_{F,i}$). We stress again how, in this regime, the Fermi energies are density-dependent, but the dependence is different for non-relativistic and ultra-relativistic particles; indeed, we have the standard expansions $E_i \simeq m_i c^2 + p_i^2/2m_i$ if $\xi \ll 1$ and $E_i \simeq p_i c$ if $\xi \gg 1$, where $p_i \simeq \pi \hbar\, n_i^{1/3}$.

From this discussion of the ideal n-p-e system, it becomes already clear how neutronization sets in. The mass difference $Q = (m_n - m_p)c^2 \simeq 1.3\,MeV$ prevents capture of low-energy electrons on protons, since there would be an energy deficit of $\Delta = Q - m_e c^2 \simeq 0.8\,MeV$. Therefore, at low densities matter in β-equilibrium is made of protons and electrons only, namely $n_n = 0$: any neutron would have decayed, while no new neutron can be formed by electron capture. However, β-equilibrium can be shifted to the neutron side whenever the electrons acquire enough kinetic energy to provide the missing $\Delta$. This is possible if matter is dense enough that $E_e > Q$; but a $1.3\,MeV$ electron is already relativistic, since its kinetic energy ($0.8\,MeV$) is larger than its rest mass energy ($0.5\,MeV$). Therefore, we expect neutronization to set in when the density is high enough that electrons become relativistic; under these conditions, not only are electrons captured by protons, but the neutrons thus produced cannot decay back, since the electron levels are already filled up to $E_e$ (Pauli blocking); the neutronized state of matter is therefore *stable*. To find the critical density for the onset of neutronization, we observe that a $1.3\,MeV$ electron has $p_e c \simeq 1.2\,MeV$ (so that $\xi_e \simeq 2.4$); this corresponds to $n_e = (p_e/\hbar)^3/3\pi^2 \simeq 7 \times 10^{30}\,cm^{-3}$. Since the number density of nucleons below threshold ($n_n = 0$) is $n = n_p = n_e$, the critical density is $\rho_\beta = n\,m_N \simeq 1.2 \times 10^7\,g\,cm^{-3}$.

The n-p-e system can be studied exactly; the β-equilibrium and charge-neutrality conditions can be rewritten as $m_p(1 + \xi_p^2)^{1/2} + m_e(1 + \xi_e^2)^{1/2} = m_n(1 + \xi_n^2)^{1/2}$ and $m_p \xi_p = m_e \xi_e$; the system of equations can then be solved numerically for $n_n/n_p$ as a function of $n = n_p + n_n$. The neutron-to-proton ratio is found to be zero for densities up to $\rho_\beta$. It then increases with increasing density, to reach a maximum value of $n_n/n_p \simeq 385$ at $\rho \simeq 7.8 \times 10^{11}\,g\,cm^{-3}$, and from there it decreases to reach its asymptotic value of $n_n/n_p \to 8$ as $\rho \to \infty$. For $\rho = 2\rho_0$ we find $n_n/n_p \simeq 130$ and $n \simeq n_n$, so that the asymmetry of typical neutron star matter is $Y_e = n_e/n = n_p/n \simeq n_p/n_n \approx 0.008$; calling these compact objects *neutron stars* is indeed quite accurate, with about one proton and one electron every hundred neutrons. Of course, so far we have neglected the strong interaction between nucleons, which can significantly affect β-equilibrium; when this is taken into account, more realistic asymmetries of order $Y_e \approx 0.05$ are obtained.

Having shown that the compact objects observed as pulsars have a maximum allowed mass of a few solar masses and that they are mostly made of gravity-compressed neutrons, we can now outline how such neutron stars are formed during the final stages of stellar evolution. We will only give the main physical ideas, and refer the interested reader to any standard book on stellar astrophysics [4]. The whole history of a star is based on an ongoing competition between gravity and pressure: as gravity forces matter to contract, different sources of pressure are exploited to counteract the inward pull and stabilize the star until its next evolutionary stage. Therefore, the initial contraction of a large and diffuse cloud of interstellar matter (H, He and traces of heavier elements) down to the formation of a protostar in quasi-hydrostatic equilibrium, which itself keeps contracting and heating up, is finally halted by the ignition in the stellar core of a thermonuclear reaction, H burning. The thermal pressure produced by fusion is then able to oppose gravity for a very long time, millions to billions of year, according to the mass (more massive stars must be very luminous to maintain hydrostatic equilibrium and thence they have much shorter lives): this phase is called the *main sequence.*

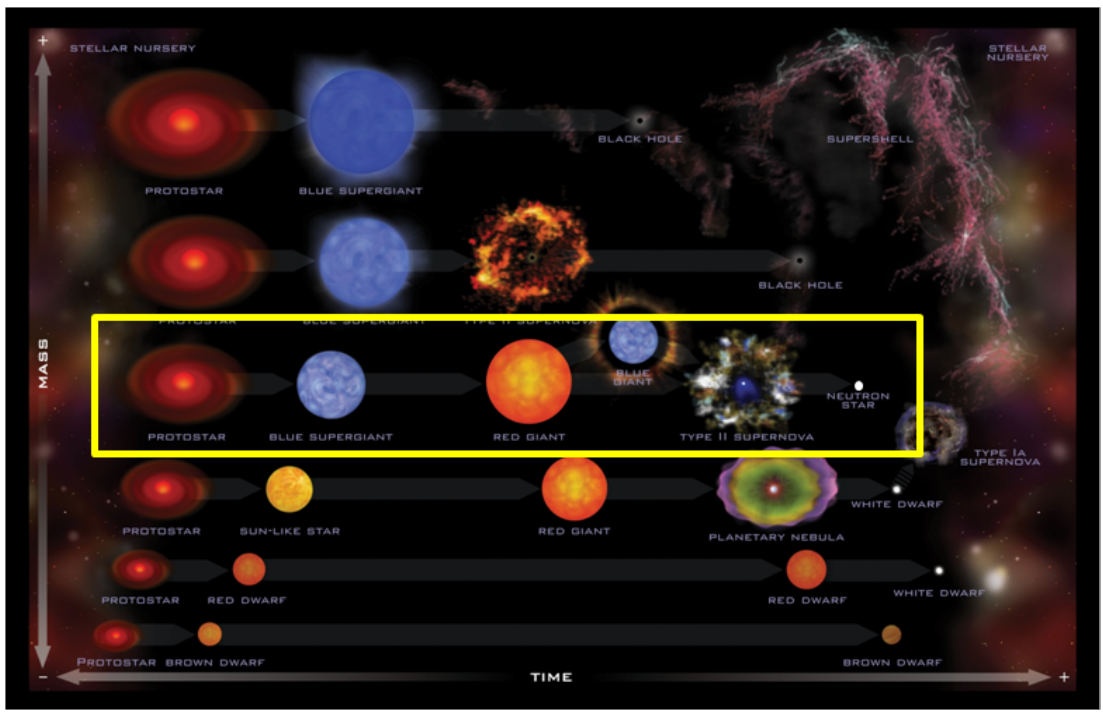

*Figure 4 – Evolutionary stages of stars as a function of the initial mass of the protostar. We highlighted the mass range that leads to neutron stars, namely $8 \lesssim M \lesssim 25\ M_\odot$, with the following stellar stages: protostar → main sequence (as a blue supergiant) → red giant → blue giant (only in some cases) → type II supernova → neutron star. Less massive stars end up into white dwarfs, more massive ones into black holes.*

When hydrogen in the core has been transformed into helium, the nuclear reactions stop and the core starts contracting again under the pull of gravity; hydrostatic equilibrium of the system requires the large hydrogen envelope that surrounds the core to expand and thence cool; the process is again halted by the ignition of nuclear reactions in the core, which burn He into C and O. The larger temperatures and densities necessary to fuse heavier nuclei with larger electric charges are thus obtained by gravitational contraction of the stellar core. The star is now a giant object with a colder surface: this is the *red giant* phase. After depletion of He in the core, the process can be repeated over and over: the core is compressed and heated up by subsequent contractions, which are temporarily halted by the fusion of the previous nuclear ashes into heavier elements (the main burning stages being H→He→C→Ne→O→Si→Fe). The extent of nuclear burning depends on the initial mass of the star, which actually determines its whole evolutionary history. This is illustrated in

Figure 4, where the main different stages of stellar evolution are shown as a function of the mass of the protostar.

Oversimplifying the actual complex phenomenology, compact stars (white dwarfs, neutron stars and black holes) represent the final, generally stable stages of stellar evolution. Low- and intermediate-mass stars with $0.1 \lesssim M \lesssim 8\ M_\odot$ can only burn H into He and, if they are massive enough ($M \gtrsim 0.5\ M_\odot$), He into C and O. The inert core then contracts and, after expelling the extended H-He envelope that surrounds it (*planetary nebula)* and by this reducing the star's total mass below the Chandrasekhar's limit ($M_{\text{Ch}} \simeq 1.44\ M_\odot$), it settles down as a degenerate *white dwarf,* which slowly cools down and gets dimmer until visual extinction. Some white dwarfs, however, can accrete matter from a companion star, if they form a binary system; when the mass exceeds the Chandrasekhar's limit, the white dwarf collapses to eventually ignite runaway thermonuclear reactions in its center; the entire star is burnt to nuclear statistical equilibrium (mostly into Fe) in a powerful explosion that completely disrupts the white dwarf, returning its matter to the interstellar medium (*type Ia supernova*). Very massive stars with $M \gtrsim 25\ M_\odot$ evolve rapidly through the main sequence and subsequent phases as hot, bloated and very luminous blue supergiants, often losing their extended envelopes already at early stages, due to strong stellar winds. At the end of their lives, they develop inert degenerate iron cores which are more massive than the mass limit for neutron stars (see Section V), so that their gravitational collapse cannot be halted by the degeneracy pressure of neutrons; either directly or after a supernova explosion, they eventually collapse into *black holes*. Predicted by General Relativity, these super-compact astrophysical objects represent the ultimate collapse of matter under the effect of gravity: the only properties left over of the original star are its total mass and angular momentum, and black holes can only be detected by the effects of their intense gravitational fields on the surrounding matter and radiation.

The stars which are expected to form neutron stars have masses in the range $8 \lesssim M \lesssim 25\ M_\odot$; they are massive enough that they can proceed through all stages of nuclear burning, all the way to Fe, but not too massive, so that the final neutron star will be below its Chandrasekhar's limit and thence gravitationally stable. Since iron is the most bound nucleus, no further energy can be obtained from thermonuclear fusion; the inert iron core, embedded in the center of the extended and massive stellar envelope, starts contracting and becomes degenerate, the zero-point pressure of electrons holding it up against gravity. Since Fe is continuously added to the core by shell-burning at its border, the core mass keeps increasing; when it reaches the value $M_{\text{Ch}} \simeq 1.44\ M_\odot$ however, it can no longer be supported by degeneracy pressure and it must collapse. The ensuing phenomenon, among the most energetic in the present Universe, is a *type II supernova*; they are also called core-collapse supernovae, to stress the physical mechanism that powers them: gravitation, not nuclear fusion [5].

Schematically, the core collapses, reaches at its center super-nuclear densities and consequently rebounds, due to the strong repulsive nature of nuclear forces at very small distances; incidentally, the physics of core collapse and rebound is also a fascinating subject for nuclear physics, since it involves *hot* and dense matter. The rebounding core creates a powerful shock wave, which violently expels the entire star envelope into the interstellar medium through an impressive electromagnetic display, the observed supernova explosion. The core itself, already very dense, hot and neutronized (proto-neutron star), contracts into its final neutron star state through the emission of a powerful burst of neutrinos, which in the case of SN1987A have been actually directly observed, thus confirming the general physical scenario. These neutrinos are also thought to play a crucial role in energizing the shock wave during its propagation across the stellar envelope. The newborn neutron star can remain in the system (like in the case of the Crab and Vela pulsars, observed in the middle of the corresponding supernova remnants), but it can also be kicked out, if the explosion is not spherically symmetric (which may be the case for SN1987A, where no pulsar has ever been detected). Notice how the core-collapse mechanism accounts for the presence of fast rotation and strong magnetic fields in young pulsars: during collapse, conservation of angular momentum and of magnetic flux can increase the original $\omega$ and $B$ of the progenitor star by factors of $\sim 10^{10}$. The gravitational binding energy of the neutron star, $\sim GM^2/R \approx 5\times10^{53}\ erg$, is mostly released in the neutrino burst: type II supernovae are actually neutrino bombs, the electromagnetic display and kinetic energy of the ejecta representing only a tiny fraction of the energy available from the gravitational collapse.

## IV – HOW EXOTIC IS THEIR INTERNAL STRUCTURE?

Having understood the origin and basic features of neutron stars, we can now turn to describe in more detail the state of matter inside them. The underlying principle has already been explained in the previous section, namely the neutronization of matter with density. In the spirit of the present lecture, we only outline the main characteristics of neutron star structure, referring the interested reader to the vast existing literature: the theoretical and experimental study of matter under extreme conditions is one of the most fascinating branches of modern nuclear physics [6].

The radial profile of neutron stars is determined by hydrostatic equilibrium under gravitational forces and by the equation of state of matter, namely its pressure and composition as a function of density. Since matter is strongly degenerate in most of the star, temperature can be neglected at first, as it will have little or no effect on the structure of a stable neutron star; temperature, however, is crucial when studying aspects like neutrino emissivity, transport of energy, cooling and so on. As a first approximation, we can take the Fermi model to represent the EOS of dense baryonic matter, namely neutrons are described as a gas of completely degenerate non-interacting fermions (like in the previous section). Even with such a simple EOS, the global features that characterize neutron stars appear clearly: for solar mass objects, Newtonian hydrostatic equilibrium yields densities ranging from zero at the surface to a few times $\rho_0$ at the center, over a radius of order $\sim 10^6 cm$. The Chandrasekhar's limiting mass, however, turns out to be quite large ( $M_{\text{Ch}} = 5.76\ M_\odot$ for pure neutron matter, the fraction of neutrons over nucleons being $Y_n = 1$), indicating the limits of such an approximation. Actually, two physical ingredients are necessary to obtain a realistic description of the star: on the one hand, an accurate EOS for matter at the different densities, which accounts for electrostatic interactions in the outer layers and for nuclear (weak and strong) interactions deeper in the star; on the other hand, General Relativity must be used to calculate the equilibrium profile (the so called TOV equation), the gravitational fields being so strong that general relativistic corrections can no longer be ignored. The details of the resulting structure depend both on the mass of the star and on the EOS used to describe its matter; while the former is to be expected (more massive objects correspond to larger central densities, since they require more pressure to be stable), the latter is due to our still limited understanding of matter under extreme conditions (we will come back to this in Section V). However, a general outline of the stratified structure of neutron stars can be given in terms of the radially increasing density.

Neutron stars are commonly divided into different regions (see Figure 5a): the *atmosphere* and the *surface*, the thin outermost layers made of *normal* matter; the *crust* (at sub-saturation densities), divided into outer and inner crust, of about $\sim 1\ km$ thickness and containing only a few percent of the total mass; the *core* (at super-saturation densities), divided into outer and inner core, of $\sim 10\ km$ thickness and including most of the star mass.

The *atmosphere*, a few centimeters thick, is where the observed thermal spectrum is formed: modern Carbon atmospheres (instead of the early H and He ones) provide excellent fits to the observed spectra and stellar radii in the range $R \sim 11.5 - 12.5\ km$. The *surface* (also called skin or envelope) is the region with density $\rho < \rho_\beta$, namely below the critical value $\rho_\beta \simeq 10^7\ g\ cm^{-3}$ for neutronization; therefore it consists of normal matter in its ground state: completely ionized $^{56}$Fe nuclei, crystallized in a Coulomb lattice, and degenerate, non-relativistic electrons providing the pressure to resist gravity. Its thermal, mechanical and electrical properties are well known, as it is the case for white dwarfs (the density range is the same, but the composition is different, since white dwarf matter has not reached nuclear equilibrium and consists of O and C); the surface presents a fascinating study in solid state physics, due to the extremely large values of density, temperature and magnetic fields as compared to laboratory conditions.

Although less exotic and much smaller than the core, the *crust* is crucial for our understanding of the physics of neutron stars. On the one hand, it represents the interface between the observable surface and the hidden core: any physical information coming from the exotic matter in the center must go through and is affected by the crust before being observed. On the other hand, some peculiar observed phenomena (for example, pulsar glitches, thermal relaxation after matter accretion, quasi periodic oscillations, anisotropic surface cooling) may originate in the crust itself. The *outer crust* is defined as the region in the density range $\rho_\beta < \rho < \rho_{\text{drip}}$ and the *inner crust* as the region in the range $\rho_{\text{drip}} < \rho < \rho_{\text{core}}$, with $\rho_{\text{drip}} \simeq 4 \times 10^{11}\ g\ cm^{-3}$ the neutron-drip density and $\rho_{\text{core}} \sim 0.7\ \rho_0$ the density where "nuclei" disappear. The equilibrium composition of matter in the crust (both below and above neutron-drip) has been extensively studied; one of the main features is the coexistence of *neutron-rich* exotic nuclei, which form a bcc (body-centered cubic) Coulomb lattice, in β-equilibrium with a strongly degenerate gas of ultra-relativistic electrons. Indeed, as explained in the previous section, increasing density induces electron capture, but this takes place on nuclei rather than on free protons and thence nuclear forces play a significant role in determining the equilibrium configuration: the nuclear species found in the outer crust have $Y_e = Z/A$ decreasing from ~0.45 ($^{62}$Ni) to ~0.31 ($^{118}$Kr) close to $\rho_{\text{drip}}$.

The other main feature is the appearance of a sea of unbound *dripped neutrons* for densities above $\rho_{\text{drip}}$: the nuclei are so neutronized that a further increase in density causes neutrons to drip out of them. Thence, in the inner crust a new phase coexists in β-equilibrium with the Coulomb lattice of nuclei (at this point better called nuclear clusters, being immersed in the neutron fluid) and the degenerate electrons. This fluid of unbound neutrons becomes increasingly important with density; at $\rho \sim 10^{12}\ g\ cm^{-3}$ it contributes about 20% to the total pressure and the electron fraction is $Y_e \sim 0.15$, while at $\rho \sim 10^{13}\ g\ cm^{-3}$ it provides already about 80% of the pressure and $Y_e \sim 0.04$. Moreover, at these densities neutrons are expected to be paired by the nuclear residual interaction (S-wave Cooper pairs); at the temperatures found in the crust, a macroscopic condensate of Cooper pairs will form and, as well known from the theory of quantum fluids, this implies that the dripped neutrons form a *superfluid*, i.e. they can flow with zero viscosity. We are in the presence of a unique scenario: a bulk, extended neutron superfluid contained by gravity in a rotating vessel (the normal, charged matter of the rapidly spinning star, namely ions, protons and electrons rigidly coupled by the strong magnetic field); we will come back to the possible observational signature of this in Section V.

At densities approaching $\rho_{\text{core}}$, nucleon-cluster structures with exotic topologies should appear, the so-called "pasta" phases ("nuclei" with rod- and plate-like shapes) whose potential relevance for observations is still unclear. For larger densities, the nuclear clusters finally dissolve into their constituent protons and neutrons: the *core* of the star is the region with $\rho > \rho_{\text{core}}$, and here we are in the presence of bulk neutron matter, the clusters having melted and neutronization being almost complete ($Y_e \sim 10^{-2}$). The neutrons in the core are expected to form a P-wave superfluid and the few protons a S-wave superconductor, the latter strongly coupled to the magnetic field; again a unique scenario of bulk (kilometer-sized) quantum fluids made of paired nucleons (the electrons are not paired in neutron star conditions and thence they form a normal fluid). At $\rho \gtrsim 2\rho_0$, muons start to appear, since at these densities the electrons have Fermi energies of the order of the muon mass: the negative charge is now carried by degenerate electrons *and* muons, in β-equilibrium with the nucleons.

The scenario we have just outlined describes the *outer core*, which will be found in all neutron stars. For low-mass neutron stars, the outer core could actually constitute the entire core, since gravity is found to compress matter in their centers by less than 2-3 times the value of saturation density. For more massive stars, however, most reasonable choices of EOS show that the density in their central regions could easily reach values up to several times saturation density. The critical densities for the appearance of new exotic phases of matter are model-dependent (according to the choice made for the potential describing the strong interaction among hadrons), so that the edge of the inner core cannot be uniquely defined. In general, one can say that for central density larger than $\rho_{\text{crit}} \sim 2-3\ \rho_0$ an *inner core* appears, which can present different exotic phases, made possible by the very large Fermi energies of the particles in the strongly compressed matter. Among the possibilities: hyperonic matter (appearance of heavy strange baryons), meson condensates (pions or kaons), *deconfined* degenerate quark matter (possibly in a state of color-superconductivity). The last one is particularly exciting, since it makes neutron stars also relevant for QCD: actually, the phase space of quark matter in the low-$T$, large-$\rho$ region can only be tested by the detection and identification of these (still hypothetical) compact objects, either entirely or only centrally composed of deconfined quarks (respectively *strange* and *hybrid* stars). We will come back to this in the next section.

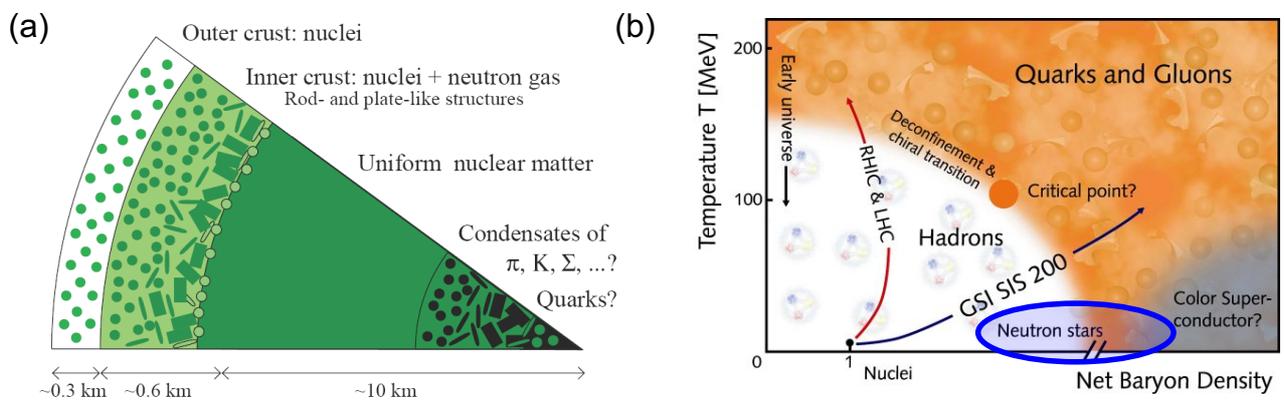

*Figure 5 – (a) Schematic profile of a neutron star. The matter composition of the different regions, their approximate spatial extensions and the transition densities between the regions are indicated (from Ref. [7-a]). – (b) Temperature-density phase diagram of strongly interacting baryonic matter. The low-T, high-$\rho$ region can only be found and studied in neutron stars.*

Of course, the fascinating microscopical scenario we just outlined is so far completely theoretical, based on our present understanding of strongly-interacting hadrons and weakly-interacting fermions compressed by gravity. The only real answer about its validity can be given by observations, in any available band: electromagnetic, neutrinos and now gravitational waves. Obviously, this is true for every stellar model at any stage of stellar evolution; in the case of compact stars, however the flow of information is almost inverted. Indeed, the microphysics of classical extended stars (the Sun being the best known example) is very well known and constrained by terrestrial experiments; at the typical densities and temperatures of the Sun, the EOS of matter is determined by the electromagnetic interaction alone. The cross sections of the nuclear reactions predicted to occur in the solar core by Hans Bethe in 1938, have now been measured directly (LUNA experiment) and the neutrinos emitted have been revealed in decades of invaluable experimental work. Future astronomical observations of the Sun, therefore, although crucial to better understand the physics of low-mass main sequence stars, are not expected to be relevant at the level of *fundamental* physics (not so long ago, however, the solar neutrino measurements have given the first evidence for neutrino oscillations, which have led to subtle modifications of the Standard Model of particle physics). The state of matter inside neutron stars, on the other hand, is neither well known theoretically nor testable experimentally on Earth: the low-temperature, high density regime determined by gravitational compression in their interiors is not reachable anywhere else in the Universe, as seen in the phase diagram of strongly interacting baryonic matter in Figure 5b. Observations of neutron stars, therefore, have the potential to constrain the still unknown properties of dense cold matter and thus to provide invaluable information about fundamental interactions in extreme conditions: in the next section we will give a few example of this fascinating investigation.

## V – WHICH OBSERVATIONS ARE RELEVANT TO FUNDAMENTAL PHYSICS?

At this point, it is clear that neutron stars represent unique objects from any directions we look at them. The question is, can we probe and study in more detail their exotic interior by astronomical observations of their external properties? Although challenging, this is actually possible and represents an exciting and active field of research involving astrophysics, nuclear physics and gravitational physics. Several and at times unexpected connections have been found between macroscopic astrophysical properties, that are already or can eventually be detected, and the microphysical state of matter hidden inside these stars. The list is long and probably useless at the introductory level of this lecture; the potential observational signatures range from mechanical to thermal to electromagnetic features, and involve sophisticated astronomical observations in all bands, as well as future detection of neutrinos and gravitational waves. Here we will just outline four significant examples, that provide a glimpse of the fascinating interplay between large- and small-scale properties of neutron stars [7].

### V.I - Mass, radius and EOS

In Section III, we discussed the Chandrasekhar's mass; the value $M_{\text{Ch}} = 5.76\, Y_e^2\, M_\odot$ (obtained for non-interacting degenerate fermions in equilibrium under Newtonian gravity) is quite reasonable for white dwarfs. Indeed, the electrostatic corrections due to the interaction between electrons and ions turn out to be quite small in these objects, and general relativistic effects are altogether negligible. The situation for neutron stars, however, is completely different: in this high-density environment, nucleons interact *strongly* through the nuclear forces; moreover, the Newtonian limit is not valid anymore in these very compact objects, and General Relativity *must* be used to study gravitational stability and determine the maximum allowed mass. As it turns out, the Chandrasekhar's mass calculated for neutron stars with the TOV equations depends significantly on the EOS used to describe baryonic matter, namely its value is determined by how we model the effect of nuclear forces. In this sense, therefore, mass measurements can yield information about and give constraints on the EOS of super-dense matter.

In Figure 6, we show a typical mass-radius diagram for neutron stars. Some regions are forbidden for gravitational stability, and this can be due to General Relativity (the star would collapse into a black hole), causality (the EOS would allow sound waves with velocity faster than light) or present rotational limits (the star would fly apart, gravity not being enough to provide the centripetal acceleration; the rotational limit is drawn for the fastest pulsar known to date, J1748-2446 with a period of 1 ms). The curves labelled with letters represent the mass-radius relation obtained upon integrating the TOV equations with different microscopic EOS: the curves to the right of the figure are for normal hadronic matter (neutron stars), those to the left for strange quark matter (strange stars). Each of these curves present a maximum, corresponding to the Chandrasekhar's mass for that EOS; it is evident that different prescriptions for the nucleon-nucleon interaction (extrapolated to higher densities from the measured properties of ordinary nuclear matter by various theoretical approaches and computational approximations), which yield different pressures and composition profiles as a function of density, predict maximum masses in the wide range $M_{\text{Ch}} \sim 1.4 - 2.8\, M_\odot$.

The potential of the mass-radius diagram is easily explained. For example, the "correct" EOS for compressed hadronic matter must be able to explain the heaviest star observed todate, namely the maximum of the corresponding curve must be higher than the horizontal line indicating the observational upper limit $M_{\text{max}}$. In Figure 6, horizontal bands indicate some of the best measured neutron star masses with their small observational uncertainties. It is clear how the new limit $M_{\text{max}} \approx 2\, M_\odot$, given by the recent observation of two heavy stars, gives strong constraints by ruling out unambiguously several theoretical models whose EOS is too soft to support a two solar mass star.

Another significant example is the suggested existence of compact stars made *completely* of deconfined quark matter (with or without a normal thin outer crust shielding it). According to a fascinating hypothesis by Edward Witten, the ground state of such a system would consist of up, down and strange quarks (this three-flavor quark matter is usually called *strange* matter) and it could be reached under extreme compression by gravity, thence the name of *strange* stars. While neutron stars (and hybrid stars, with quark matter only in their inner cores) are bound by gravity, and their radius decreases for increasing mass, strange stars are bound by gluons, and their radius increases with mass. For the same mass, strange stars are more compact than neutron stars and thus they are able to sustain faster rotations. For example, a one solar mass pulsar with a period of half a millisecond would unequivocally identify it as a strange star since, in order to satisfy the Keplerian limit ($\Omega < \Omega_k = \sqrt{GM/R^3}$), such an object should have a radius of less than ten kilometers: from Figure 6, we see that this is only consistent with a quark matter EOS. More generally, future observation of a half-millisecond pulsar would expand significantly the shaded region forbidden by rotational stability, leaving only strange stars above the new rotational limit. Of course, such an observation would not rule out standard neutron stars but would be unequivocal evidence of the *existence* of a new family of compact objects. As for the already known pulsars, to date they are consistent with neutron or hybrid stars: the "smoking gun" for the existence of strange star is still missing.

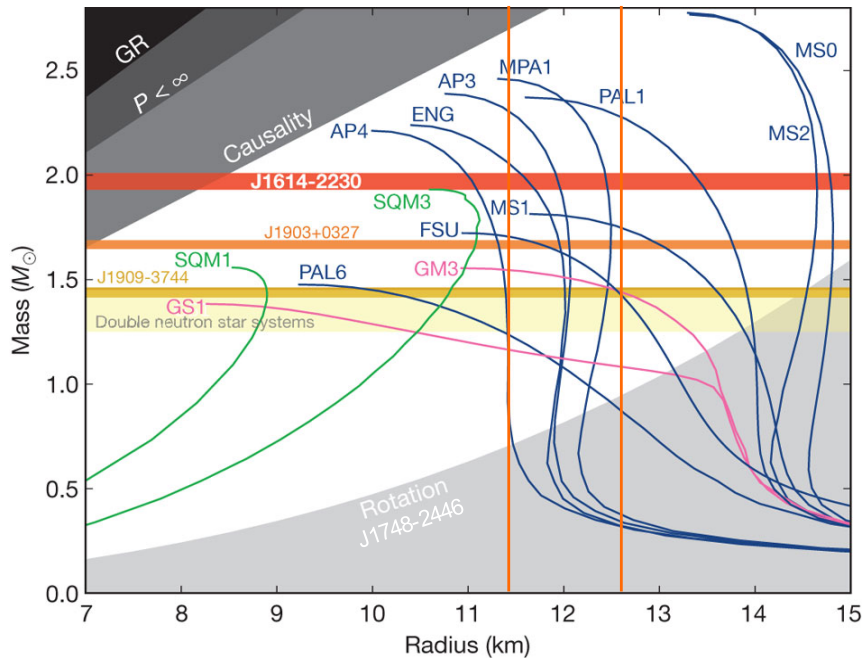

*Figure 6 – Mass-radius relation for different EOS of dense matter. The curves to the right of the figure are for normal hadronic matter, those to the left (labelled SQM) for strange quark matter. Regions of the diagram excluded by rotation, causality and General Relativity are also given, while the masses (with their observational uncertainties) measured for different neutron stars are indicated by horizontal bands (from Ref. [3-a]). The range $R \sim 11.5 - 12.5 \, km$ for the radius is delimited by vertical lines.*

The best case, of course, would be a direct measurement of mass *and* radius for the same star: instead of just constraining the possible classes of EOS, this would precisely indicate a point in the diagram where the theoretical mass-radius curve must pass, thus picking out a particular EOS out of the several which have been constructed, using different prescriptions for the nuclear forces or different approximations to describe the extended system of strongly interacting nucleons (or quarks for the deconfined phase). So far it has not been possible to determine mass and radius for the same neutron star with enough precision: as mentioned, radius measurements are still an open issue and although it can further constrain the EOS, the range $R \sim 11.5 - 12.5 \, km$ is still compatible with several existing nuclear models. However, the possibility in the future of such an observation would be invaluable to nuclear physics, providing a big step forward in our understanding of the nature of nuclear forces.

A last word of caution: the diagram in Figure 6 seems to suggest that the theoretical scenario is under control and that observations can be explained by at least some of the models developed so far to describe the nucleon-nucleon interaction. There is, however, a serious problem with *hyperons*, which were not included in any of the displayed EOS; the interaction cross sections of these heavy strange baryons with nucleons are poorly constrained by accelerator experiments. Using the presently known accepted values the following scenario seems unavoidable: hyperons are predicted to be present at the typical densities of heavy neutron star cores, but their inclusion softens dramatically the EOS of matter so that the maximum calculated mass is *always* smaller than two solar masses. At present, therefore, our microphysical models are contradicted by observations: this "hyperon puzzle" must be solved before we claim to understand strongly interacting dense matter. The answer may be found in better cross section measurements; or, like other puzzles before (the solar neutrino puzzle making a good case), its solution may have to invoke new fundamental physics.

### V.II – Cooling of neutron stars

The launch of dedicated X-ray satellites in the last decades has made it possible to study the surface temperature of several isolated neutron stars. When coupled with an estimate of the age of the star, these observations provide an alternative way to investigate the properties of the exotic interior. Very dense and/or very hot matter cools mainly by direct neutrino emission (*lepton cooling*), rather than through the transport of heat by photons or electrons (radiative and conductive cooling). According to the physical conditions, neutrinos can be produced by different processes, thermal (neutrino-pair production) and nuclear (electron capture). Neutrino cooling is crucial in the advanced stages of evolution of massive stars, in the gravitational collapse of their iron core, in the evolution of the proto-neutron star as it contracts into its stable compact configuration, and in the cooling of the young neutron star. Neutrino emission is an extremely efficient cooling mechanism, which draws energy directly from the dense core; neutrinos emitted in the centre of a neutron star, with typical temperatures $T \lesssim 10^9 \, K$, escape freely out into space without interacting with matter, as opposed to the more common radiative or conductive processes, where heat carried by photons or electrons diffuse slowly from the centre to the surface, transferring energy which will eventually be radiated into space. Incidentally, during the very hot ($T \sim 10^{10} - 10^{12} \, K$) proto-neutron star stage even neutrinos significantly interact with matter (mostly with nucleon, by neutral-current scattering processes and by charged-current neutrino absorption) and thence diffuse on timescales of seconds, as observed directly in the neutrino burst from SN1978A: this is quite exceptional for these elusive particles and provides a powerful insight into the physics of weak interactions.

The result is that the crust of a young neutron star, where slow conductive cooling dominates (heat transfer is mostly conductive when degenerate electrons are present), is soon much hotter than the core, which cools by lepton emission. The subsequent thermalization of the star, where heat diffuses *inward* from the crust to the core, determines the cooling history of the observable radiating surface. Seen from another angle, we can say that the thermal signature from the cold interior must diffuse through the insulating hotter crust around it, before reaching the surface and being detectable by an outside observer. The purpose of cooling studies is to calculate by detailed numerical simulations the observed surface temperatures and luminosities of neutron stars as they age; theoretical cooling profiles obtained for different physical input parameters are then compared to observations. Several important factors determine the shape of the cooling curves: stellar mass,

EOS, neutrino emissivities, specific heats, thermal conductivities, pairing properties of matter, magnetic fields; therefore, a systematic analysis can be carried on, to confirm or rule out some of the many possible scenarios and thus test indirectly the state of matter inside neutron stars. Next, we give two examples of how cooling is affected by the microscopic state of matter and can thus be used as a diagnostic tool.

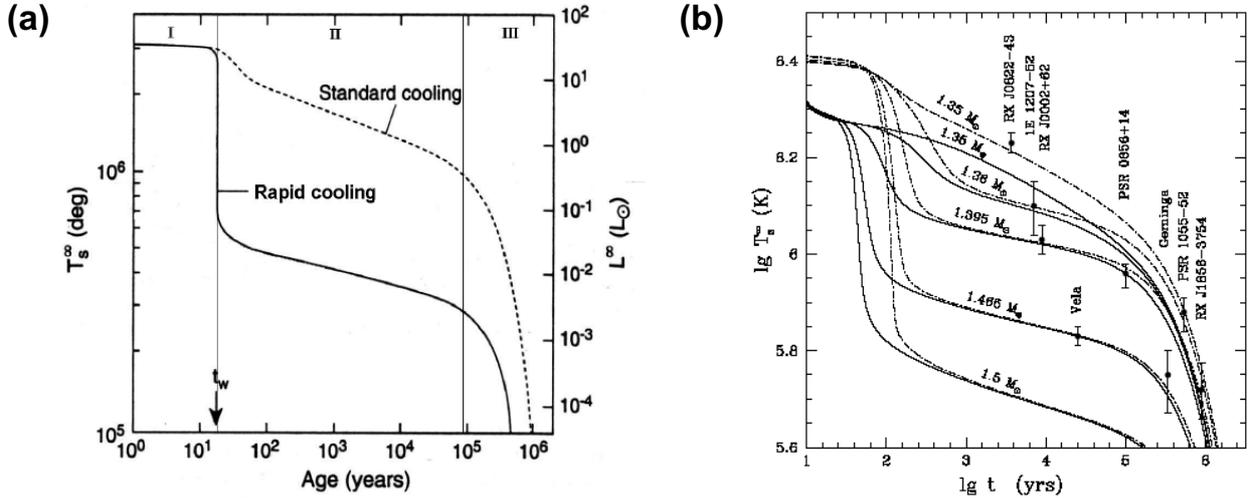

Figure 7 – (a) Typical cooling curves for standard (dashed) and rapid (solid) cooling of the core. The three vertical regions correspond to: I) core relaxation epoch; II) neutrino cooling epoch; III) photon cooling epoch (from Ref. [8]). – (b) Example of cooling curves calculated for different neutron star masses and with (solid) or without (dashed) neutron superfluidity in the inner crust; observations are also indicated with their error bars (from Ref. [3-d]).

In Figure 7a, we show two typical theoretical cooling curves: the surface temperature (left scale) and luminosity (right scale), as observed by a distant observer (at infinity), are given as a function of the age of the neutron star. Three temporal regions can be individuated, which characterize different cooling phases: the core relaxation epoch (I), the neutrino cooling epoch (II), and the photon cooling epoch (III). During core relaxation, the surface temperature is practically constant; although the core has already cooled by neutrino emission and become isothermal, the thermal inertia of the surrounding crust still prevents the thermal signal to reach the surface and affect its temperature; the length of the plateau phase (in the range $10^1 - 10^2$ $yr$ after formation) depends on the thermal and pairing properties of the crust. During epoch II, the surface responds to the cooling of the deep interior and its temperature decreases; the overall loss of energy from the star is still dominated by neutrino emission from its denser parts. After $\sim 10^5$ $yr$, the interior of the star has cooled enough that neutrino emission becomes inefficient: during epoch III, the cooling of the star is determined by emission of photons from the surface, with standard conductive heat transfer inside; the different cooling mechanism explains the different slope of the curves in regions II and III.

The two curves in Figure 7a refer to two different types of neutrino cooling in the stellar core: *standard cooling* processes are slower and take place in standard neutron matter, namely in the outer core; *rapid cooling* processes are very fast and they are associated to the different exotic states of matter (meson condensates, hyperonic matter or quark matter), which appear if the central density is large enough for the formation on an inner core. Altogether, standard cooling would indicate low-mass stars with a stiff EOS (so that the central densities remain below the threshold for the appearance of exotic matter), while rapid cooling would be typical of either low-mass stars with a soft EOS or high-mass neutron stars (dense enough to present an exotic inner core). As a product of the supernova explosion, the core is formed at $T \sim 10^{11} - 10^{12}$ $K$ ; it is so hot that matter is only weakly degenerate and it cools to temperatures $T \sim 10^9$ $K$ in a few days, by intense neutrino emission (the so-called *direct-URCA* process, namely energy draining cycles of electron captures and $\beta$-decays). When $T \lesssim 10^9 \, K$, matter in the core becomes degenerate and the rate of neutrino cooling is determined by the composition of matter, which in turn depends on the density reached under gravitational compression. For normal neutron matter, only standard cooling (the so called *modified-URCA* process, where a spectator nucleon must enter the weak reactions to conserve energy and momentum thus lowering the cross section) is allowed; in this regime, the core cools to $T \sim 10^8 \, K$ in a few centuries. For exotic matter, instead, fast cooling (analogous to the direct-URCA) is permitted and the core can reach temperatures of $T \sim 10^8 \, K$ in just a few decades.

The observational signature for the existence of an inner core of exotic matter is evident from Figure 7a: a sudden drop in surface temperature of a young neutron star would be unequivocal evidence of the presence of such a core. Moreover, older stars which are too cold as compared to stars of similar age and thence are not fitted by standard cooling (e.g., the Vela pulsar in Figure 3a) could also indicate the presence of exotic matter in their interiors; on the other hand, cooling calculations have ruled out the Vela as a possible candidate for a strange star with crust. Incidentally, the time of rapid cooling, $t_w$, if ever observed could also indirectly constrain the EOS of dense matter [8].

As a second example, we consider Figure 7b; here, two sets of cooling curves are obtained for different neutron star masses from cooling calculations *with* (solid lines) or *without* (dashed lines) neutron superfluidity in the inner crust. Nucleon pairing affects cooling in two ways: it quenches neutrino emissivities and lowers the specific heat of superfluid nucleons. Several general features can be noticed: the surface temperatures are quite sensitive to the mass, thus suggesting an alternative way of measuring this quantity (e.g. for isolated stars); pairing is relevant only for young neutron stars (few centuries old) unless they are very light; rapid cooling sets in when the star is heavy enough to develop an inner core; the Vela and Geminga should be more massive than the other observed pulsars. Of course the problem of fitting individual stars and thus constrain microphysics is highly degenerate: the different input (mass, EOS, pairing, onset of rapid cooling, magnetic field and so on) act simultaneously in non-linear ways. The general philosophy, however, is clear: the thermal evolution of the surface is seen to be quite sensitive to the properties of matter, both in the core and in the inner crust, and cooling can be used to probe indirectly these properties.

As a last remark, we come back to the case of the CCO in Cassiopeia A. As already mentioned, the beautiful data extracted from 15 years of observations by the *Chandra* X-ray satellite are the first direct evidence of neutrino cooling of a given star with time, allowing to check our models for neutrino processes. In particular, the faster rate of decrease in *temperature* (inset of Figure 3b) around 300 years after birth has been interpreted as evidence for superfluidity (in this case, the onset of the breaking and formation of neutron Cooper pairs in the $^3P_2$ channel) in the star interior: another example of the fascinating interplay between fundamental physics and observations of neutron stars [3-d].

### V.III - Pulsar glitches

Pulsars show a regular slow-down in their rotational frequency due to the emission of energy at the expenses of the rotational kinetic energy, as explained before; also magnetars and thermal emitters are slowed down by electromagnetic torques, although this is only a fraction of their luminosity which is mostly sustained by magnetic and residual thermal energy respectively. Several pulsars, however, present sudden spin ups called *glitches* (see Figure 8a); the Vela pulsar, for example, has been glitching once every 2-4 years for the last thirty years, with jumps of the order of $\Delta\Omega \sim 10^{-6} \, \Omega$ and $\Delta\dot{\Omega} \sim 10^{-2} \, \dot{\Omega}$; also magnetars and old millisecond pulsars have been observed to glitch. To explain what appears as a quite common phenomenon, which involves a massive and sudden transfer of angular momentum to the observable crust, several models have been proposed with different internal mechanisms and reservoirs to store for a few years and then rapidly release to the crust the angular momentum needed to cause a glitch. On the one hand, the spin-up of the crust has not yet been resolved and the present observational upper limit to the risetime is *less* than one minute: this implies that the glitch itself is an almost instantaneous, catastrophic event. On the other hand, the long timescales of the exponential-like *recovery* observed after large glitches (of the order of weeks or months) can only be explained by the existence in the star of at least two mutually-interacting components: a normal rigid crust very weakly coupled to a *superfluid* in the interior; indeed, a normal fluid would interact strongly with the crust, resulting in relaxation times of order of minutes. With this kind of phenomenological model, the observed after-glitch recoveries can be well fitted and reasonable constraints can be put on the weak coupling between the normal and superfluid components.

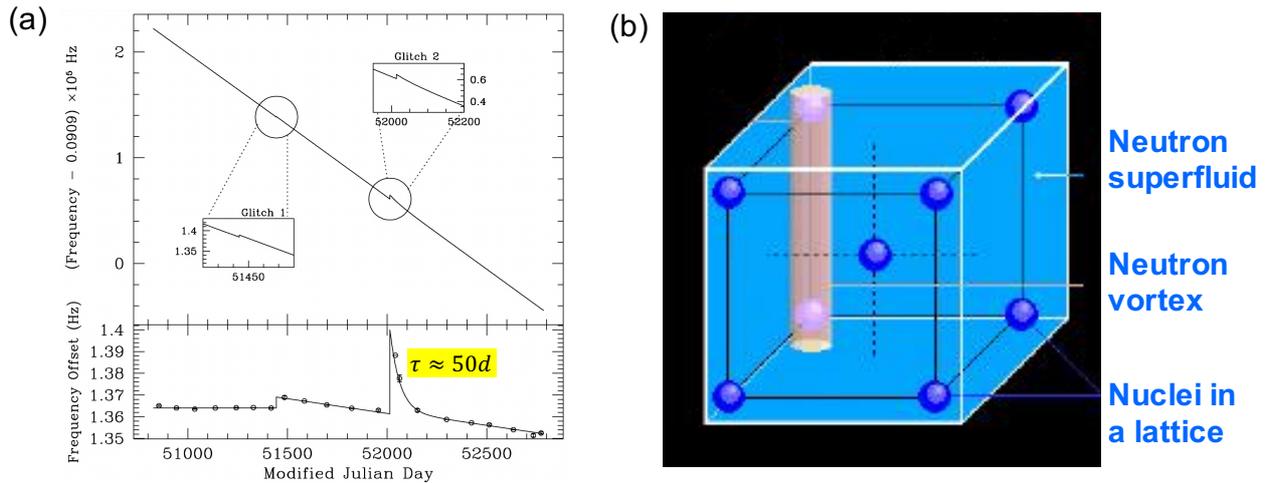

*Figure 8 – (a) Observations of glitches in the decreasing rotational frequency of a neutron star; in the lower panel, the secular spin down has been subtracted. The second larger glitch exhibits the typical exponential-like recovery, with long timescales of order $\tau \approx 50\ days$. – (b) Interaction between a vortex in the neutron superfluid and the lattice of exotic nuclei in the inner crust of neutron stars.*

The association of glitches to superfluidity, however, goes further: one of the most promising and intriguing explanation of pulsar glitches is the so-called *vortex model*, which is based on the coexistence of a neutron superfluid and a lattice of nuclei in the inner crust of neutron stars. In this model, the *reservoir* of angular momentum is the neutron superfluid itself; to date, this is the only viable explanation for the *giant* glitches ($\Delta\Omega \sim 10^{-4}\ Hz$) observed repeatedly in several pulsars (e.g., the Vela). The mechanism invokes a remarkable property of rotating superfluids, which is well known and observed in the laboratory for superfluid $^4$He and $^3$He: because of the presence of a quantum condensate, superfluid flow must be irrotational and thus a superfluid cannot rotate as a rigid body, since rigid rotations have non-zero curl (if $\vec{v} = \vec{\Omega} \times \vec{r}$ then $\vec{\nabla} \times \vec{v} = 2\vec{\Omega} \neq 0$). Superfluids in a rotating vessel, instead, develop an array of quantized vortex lines which are parallel to the rotational axis of the container: these represent singularities in the rotational flow (a line vortex has a velocity profile $v(R) \propto 1/R$ where $R$ is the distance from the vortex axis) and they carry quantized angular momentum. Vortices are like excitations of macroscopic energy and angular momentum in the superfluid, which locally (in the vortex core) destroy superfluidity and thus allow a coupling of the superfluid to normal matter (e.g., electrons can scatter off the vortex cores). The dynamics of the superfluid must then be described in terms of distribution and motion of vortices, and an *average* macroscopic superfluid velocity can be defined in terms of the density of vortices. The superfluid can decrease its angular momentum only by expelling vortex lines from its interior; if for some reason (e.g., impurities) the lines cannot move, then the average velocity of the superfluid cannot change so that its angular momentum and rotational pattern are *frozen*.
As discussed in Section IV and shown in Figure 8b, the inner crust of neutron stars is characterized by a superfluid of unbound neutrons permeating a Coulomb lattice of exotic nuclei

(clusters of bound nucleons). The normal matter of the star (nuclei and electrons, coupled by the strong magnetic field throughout the star) forms a rigidly rotating container, so that the neutron superfluid in the inner crust develops an array of vortices, whose number density determines the angular velocity of the star (Feynman-Onsager relation [9]). The nuclear lattice, instead, represents a system of impurities (the nuclear clusters) to which the vortices can potentially pin and thence lose their mobility. In the case of strong *pinning* of the vortex lines to the lattice, the following situation develops: the crust and the vortices pinned to it slow down as the star loses rotational energy, but the neutron superfluid is frozen and its average velocity cannot follow the slow down. An increasing velocity lag between the superfluid and its vortices then develops, and in turn this applies strong hydrodynamical lift forces (Magnus force) on the vortex lines, which tend to unpin them from the lattice of nuclei. When the Magnus force, which is proportional to the velocity lag and thence increases with time, becomes larger than the pinning force which binds the vortex lines to the lattice, the vortices catastrophically (like an avalanche) unpin from the lattice and are then able to transfer their angular momentum to the normal matter of the outer crust, thus spinning it up and causing a glitch. Once unpinned, the vortices still interact with normal matter through various dissipative drag forces which exchange angular momentum: the known mechanisms predict risetimes of order of seconds and slow recoveries, as actually observed.

The stratified structure of neutron stars implies differential rotation of the neutron superfluid, so that the simple model with two rigid components is only a first rough approximation; subtle phenomena, like superfluid entrainment (a non-dissipative interaction between neutron and proton mass currents, which can reduce the angular momentum reservoir) or vortex creep (thermally activated unpinning of vortices through quantum tunnelling), must be taken into account. We do not even try to go into any detail, the approach being quite complex and requiring specific knowledge of vortex dynamics and of their interaction with normal matter (pinning and drag). Although very promising, since it provides a sufficient reservoir of angular momentum to explain large glitches, the vortex model has not yet proven to be the correct explanation, also because some fundamental questions are still open. First of all, the *trigger* that initiates a glitch has not yet been identified among the different mechanisms (e.g., starquakes, hydrodynamical instabilities, self-organized criticality). Then, neutron vortices are also present in the core of the star, and here they interact with the magnetic flux tubes associated with proton superconductivity: this is still a quite unchartered issue. Even more crucial, the pinning force that binds the vortex lines to the lattice of exotic nuclei is still poorly known: the interaction of a vortex with a *single* nucleus in the neutron-drip regime is already a hard problem to treat quantistically, and extending the calculation to the whole lattice is even more forbidding. Still at the microscopic level, the pairing properties of nucleons in dense matter are also an open issue, this being part of the more general theoretical uncertainties concerning the nucleon-nucleon interaction which we already mentioned.

Reasonable approximations, however, are also possible and the results so far are not inconsistent with observations. The main point we want to stress here is the following: if the vortex model is the correct explanation, then pulsar glitches would represent unique and direct macroscopic *evidence* of the existence inside neutron stars of an extended (kilometre-sized) neutron superfluid, which interacts significantly with the nuclear lattice via its vortex lines. The idea is fascinating and allows an independent way to probe some nuclear properties of matter (e.g., neutron pairing and structure of exotic nuclei beyond neutron-drip) inside neutron stars.

### V.IV – Compact stars and the structure of space-time

General Relativity (GR), the theory of curved space-time proposed by Albert Einstein to explain gravitation, is usually associated to cosmology and its global description of the Universe; its main application to stellar physics seems to be the study of black holes, which can only be understood in this framework. Neutron stars and the systems they form, however, also require GR instead of Newtonian gravity to describe them and this is due to their extreme *compactness* (mass to radius ratio) which implies strong gravitational fields. To see this one can consider a related quantity, the adimensional potential $\Phi = GM/Rc^2$ which is also the ratio between gravitational binding energy

($\sim GM^2/R$) and rest mass energy ($Mc^2$) of a star. In the Newtonian theory, the source of the gravitational field is the rest mass alone, while in GR it is the total mass-energy; therefore, the Newtonian limit must correspond to a negligible contribution from the gravitational energy, namely to $\Phi \ll 1$. Conversely, the ultimate general relativistic object are *black holes*, where matter has collapsed into a central singularity; the Schwarzschild's radius $R_S = 2GM/c^2$ corresponds to the event horizon, namely the boundary of the region of space-time that cannot communicate with the external universe and that delimits the black hole itself; therefore the fully GR limit corresponds to $\Phi = 1/2$. Classical stars are extended objects and thence can be described by Newtonian hydrostatic equilibrium (for example, the Sun has $\Phi \approx 10^{-6}$); the compactness of neutron stars, however, is such that $\Phi = 0.2 - 0.3$ so that GR *must* be used (the TOV equations, as already mentioned). Moreover, the surface gravitational potential is so large that photons emitted with energy $h\nu_0$ are significantly redshifted when escaping the gravitational well of the star and are observed at infinity with a smaller frequency $\nu < \nu_0$; conservation of energy gives the *gravitational redshift* formula $|\Delta\nu|/\nu_0 = \Phi$. Directly related to this, in Figures 3 and 7 the blackbody surface temperatures and luminosities obtained from cooling calculations are suitably redshifted for a distant observer.

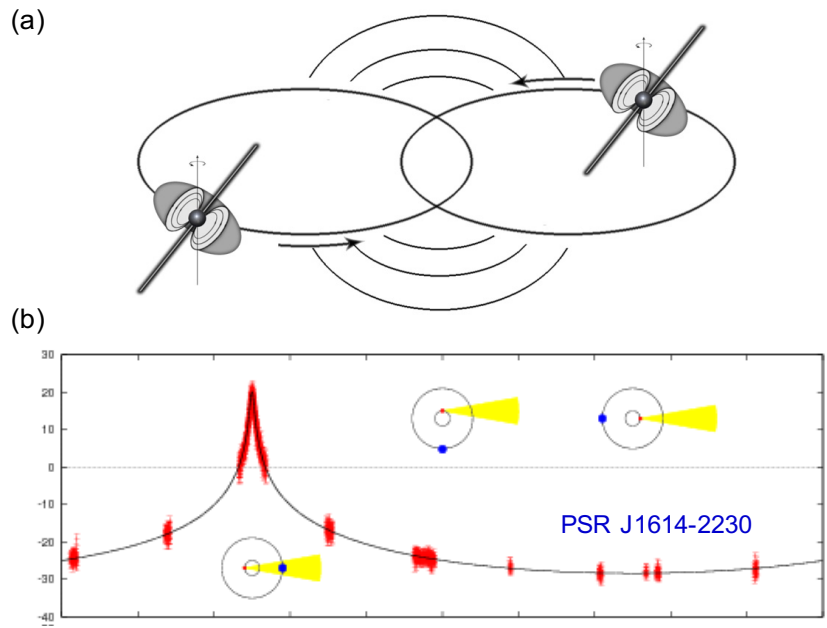

*Figure 9 – (a) Binary system of two pulsars: the ultimate lab for General Relativity. – (b) Mass measurement of the millisecond pulsar PSR J1614-2230 using general relativistic Shapiro delay. The pulsar's signal is delayed when crossing the curved space-time around the companion white dwarf, allowing to infer the masses of both stars (from Ref. [3-e]).*

If isolated neutron stars already require GR to determine their gravitational hydrostatic equilibrium, it is actually *binary* systems containing one millisecond pulsar and another compact object (white dwarf, black hole or a second neutron star, like in Figure 9a) that are particularly relevant to investigate the properties of space-time. Indeed, the intense gravitational fields associated to these compact systems allow unique studies of gravity in the *strong-field* limit (the classical tests of GR in the Solar system, instead, are all in the weak-field limit): although the orbital potential is still small ($\Phi_{orb} \approx 10^{-6}$), if GR is not the correct theory the presence in the system of a neutron star with $\Phi_{NS} \approx 10^{-1}$ can be revealed in the orbital motion. Moreover, the fast millisecond pulsar provides a high precision clock (better than atomic clocks) crucial to detect the subtle effects to the orbital motion due to the curvature of space-time and compare them to the predictions of GR. Last but not least, these effects allow very accurate mass measurements, otherwise impossible.

As a first example, in Figure 9b we illustrate the mass determination for the millisecond pulsar PSR

J1614-2230, one of the two massive neutron stars mentioned before. The binary companion is a white dwarf and the orbit happens to be edge-on with respect to Earth, so that the *Shapiro delay* could be observed: the travel time of the pulsed signal is longer when the pulsar is behind the white dwarf, due to the curved space-time around the latter which the signal must cross. This GR effect coupled to Kepler's law yields the value $1.97 \mp 0.04\ M_\odot$ for the neutron star mass [3-e].

As a second example, we consider the Double Pulsar (PSR J0737–3039); this is the ultimate lab to study GR, since the two pulsars can be viewed as test masses each with a precision clock attached to it. The system is extremely compact, with an orbital period of 147 minutes, and it exhibits several measurable GR effects: precession of periastron ($\sim 10^5$ times that of Mercury in the Solar system), gravitational redshift, Shapiro delay, geodetic precession [7-b]. These effects can be expressed by the so called post-Keplerian parameters, which depend on the observed orbital parameters and on the two masses. When plotted in a mass-mass diagram using the equations of GR, each effect determines a strip due to observational error on the corresponding parameter, as shown in Figure 10a. If GR is the correct theory of gravity the lines corresponding to the post-Keplerian parameters should meet in a single point; the fact that these strips simultaneously intersect each other (blue region in the inset) constitutes a stringent test of GR within the observational uncertainties (alternative theories of gravity should perform at least as well as this). Moreover, the small region of intersection allows a determination of the two masses with an incredible precision of better than 0.1%; this is unheard of in stellar astronomy and is made possible by the extreme accuracy of pulsar timing.

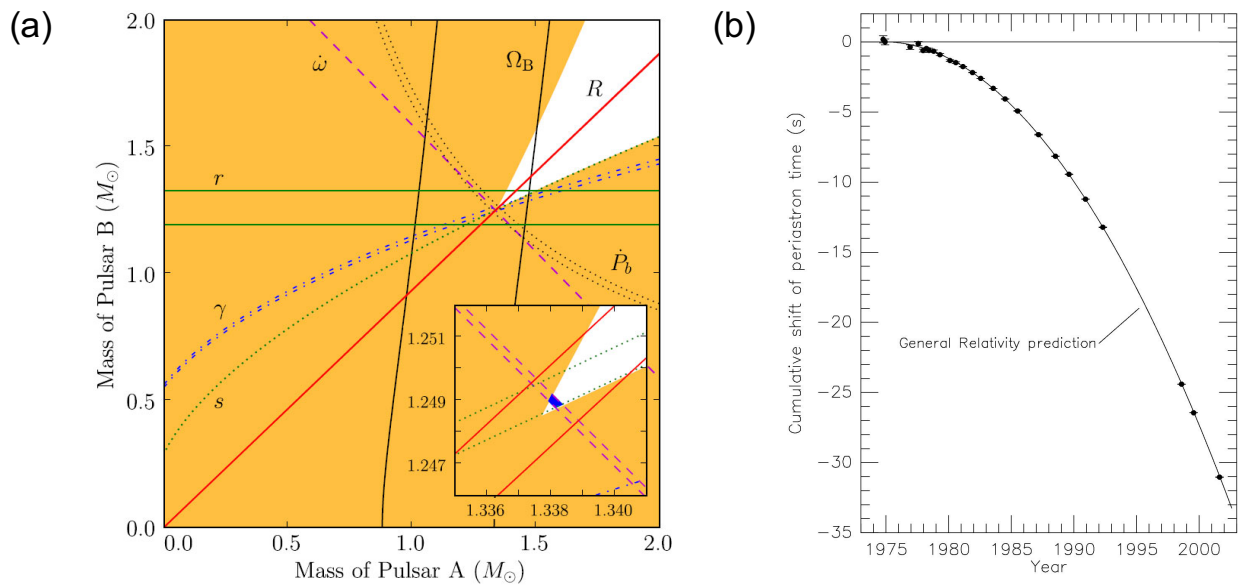

*Figure 10 – (a) Mass determination and test of General Relativity in the binary system PSR J0737–3039 (Double Pulsar): observations of different post-Keplerian orbital parameters coupled to GR delimit strips in the massA-massB diagram. Their intersection provides a stringent test of GR and gives a high-precision measurement of both masses. – (b) Evidence for gravitational wave emission from the Hulse-Taylor binary pulsar: thirty years of observations of the orbital period decay (dots) are compared to the prediction of General Relativity (both figures from Ref. [7-b]).*

Gravitational waves, namely tiny ripples in the metric of space-time propagating at the speed of light, are among the most intriguing predictions of GR. Detecting them has been one of the greatest scientific challenges and again compact objects are playing the main role in this fascinating investigation. Binary systems of compact stars have a time-varying mass quadrupole and are thus expected to be continuous sources of gravitational waves; the associated loss of energy results in a decay of the orbital motion (the orbital radius shrinks and the orbital period decreases). In Figure 10b, thirty years of observation of the Hulse-Taylor binary pulsar (the first one discovered) are reported: the observed orbital period decay is indicated by the points and the predictions of GR by the line. The agreement is impressive and constitutes the first indirect *evidence* for the existence of gravitational waves; similar results hold for the Double Pulsar and the other known binary pulsars.

The loss of energy by emission of gravitational waves and the associated orbital decay imply that eventually the two compact stars in a binary system must coalesce and merge into a black hole (their total mass usually exceeding the Chandrasekhar's limit); this cataclysmic process is expected to produce an intense burst of gravitational waves. On September 14, 2015 the first ever observation of such a merger at the two *Ligo* detectors [10] has finally given direct proof of their existence (Figure 11a). The sheer power of this event is breathtaking: about $3M_\odot$ of energy ($\sim 10^{54}\ erg$) were emitted in gravitational waves, more than ten times the energy emitted in neutrinos by SN1987A; the merger involved two massive black holes (a study case of pure GR, in the strong-field regime), namely a scenario where no electromagnetic counterpart is expected and whose *only* signature is a burst of gravitational waves. This has marked the beginning of a new era by opening an alternative observational window into the physics of compact stars and their final evolutionary stages.

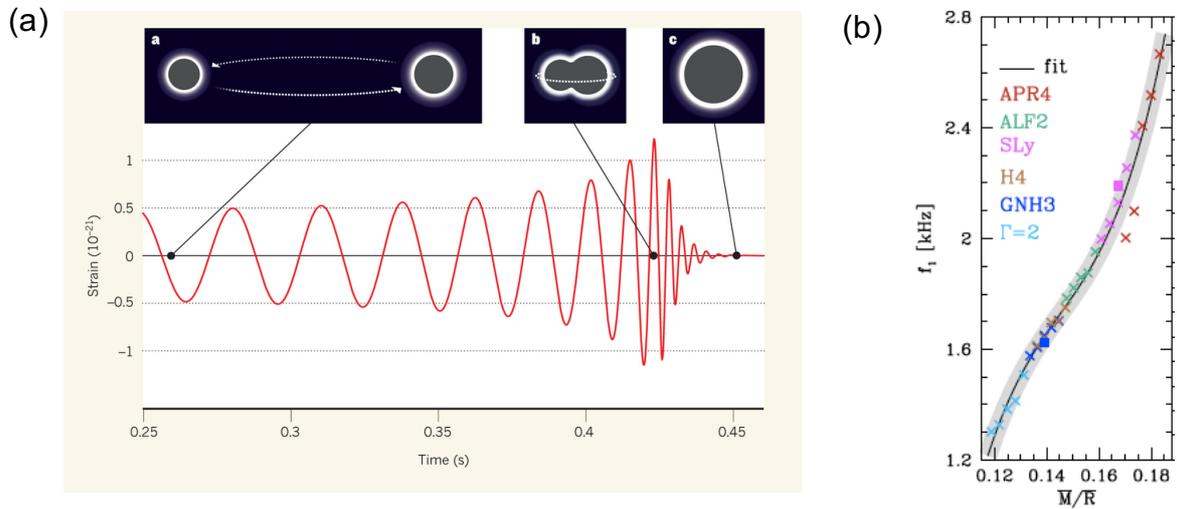

*Figure 11 – (a) The first gravitational wave detection: GW150914, the merger of two massive black holes. The three phases of the event (inspiral, merger and ring-down to the final black hole) are indicated above the gravitational wave signal (from Ref. [10]). – (b) 'Universal' correlation between the low-frequency peak in the power spectral density of the postmerger gravitational wave signal and the 'average compactness' of the two coalescing neutron stars (adapted from Ref. [11]).*

Although the first observed merger involved two massive black holes, the *Ligo* and *Virgo* detectors are now sensitive enough to reveal the merger of two neutron stars, an event whose electromagnetic counterpart is usually associated to short gamma-ray bursts. The gravitational wave signal produced by this merger is expected to carry invaluable information about the properties of dense matter: the imprint of the EOS on the waveform is a thriving area of research in numerical relativity, providing a new tool to investigate hadronic interactions. For example, detailed numerical simulations in full GR have shown a remarkable correlation between the low-frequency peak in the postmerger emission spectrum and the 'average compactess' of the two stars; as shown in Figure 11b, the correlation is *universal* (independent from the EOS describing dense matter). Another correlation (this time EOS-dependent) with the 'average mass density' has been found for the second peak in the power spectral density: together, these relations would enable to fix a point in the M-R diagram of Figure 6 already after a single detection of the merger of two neutron stars, thus selecting one EOS over the others [11].

Gravitational waves probe global, large-scale properties of self-gravitating systems and travel almost unhindered across space; in this respect they are orthogonal to electromagnetic waves, which depend on the small-scale details of the source and are significantly affected by the interstellar medium in their propagation. These two observational bands are thus *complementary* tools in our investigation of the fundamental properties of matter and space-time using neutron stars: new exciting discoveries are right behind the corner.

## VI – CONCLUSION

We have reached the end of this lecture. Although only a first introduction to the physics of neutron stars, it shows how the main properties of these compact objects follow directly from general physical principles: when matter is compressed by gravity to its extreme limits (i.e., short of becoming a black hole), which can happen only as the result of the long evolutionary process of massive stars, the final object naturally shows extreme physical conditions, otherwise unreachable. Better understanding of neutron stars can only come from an interplay of multi-band observations, astrophysics, nuclear and gravitational physics. Since observations of macroscopic features of the star must be explained in terms of the microphysics in their interiors, a strong interplay of theorists from different areas of physics is necessary: the European network *Compstar* is an example of such an interdisciplinary approach [2].

From the point of view of observations, the beginning of the new millennium has been a particularly exciting time; satellite-based observatories in different bands are giving our scientific community an incredible wealth of new data: several new classes of neutron stars have been discovered in the past decade and the results from the *Fermi* $\gamma$-telescope have changed some of our views about these objects. We expect the new generation of satellites and ground-based radio-telescope arrays to reserve further surprises. Moreover, the now fully operational detectors of gravitational waves (e.g., *Virgo* in Italy and *Ligo* in the US) will soon open a new window for the investigation of compact objects by observing the merger of two neutron stars (astro-seismology by gravitational waves), while large neutrino detectors (e.g., *SuperKamiokande*) make the possibility of extra-solar neutrino astronomy (started with SN1987A) a concrete option. Finally, heavy-ion colliders (e.g., *RHIC* and now *LHC)* will enable nuclear scientists to probe some properties of exotic nuclei; although the conditions found inside neutron stars (even those in the inner crust) are well beyond the reach of any experimental facility, direct data about the isospin-asymmetric matter formed in heavy-ion collisions could help nuclear theorists to understand whether they are on the right track.

Only from a synergic effort of all these different fields of expertise we will be able to reach a coherent description of neutron stars: a unique environment in our present Universe, where gravitation dominates over the other fundamental interactions and, by compressing matter to its limits, determines extreme physical scales and exotic scenarios, both at the macroscopic and microscopic level. Few fields of physics are as inter-disciplinary as the study of neutron stars, and present so many challenges but also so many promises to reveal new fundamental physics or test existing theories.

## ACKNOWLEDGEMENTS

This lecture was given at the Ecole Joliot Curie 2016, in the nice atmosphere created by the organizers and participants of the school at Port-Barcarès. Some of the figures are taken from the web. Partial support comes from 'NewCompStar', COSTAction MP1304.